\documentclass[12pt]{article}
\usepackage{eurosym}
\usepackage{float}
\usepackage{graphicx}
\usepackage[utf8]{inputenc}
\usepackage[T1]{fontenc}
\usepackage{indentfirst}
\usepackage[margin=0.6in,nomarginpar]{geometry}
\usepackage[final]{hyperref}
\usepackage{amsmath}
\usepackage{hyperref}
\usepackage{cite}
\usepackage{subcaption}
\usepackage{caption}
\usepackage{amssymb}
\usepackage{multirow}
\usepackage[table]{xcolor}
\usepackage{orcidlink}
\usepackage{color}
\usepackage{hyperref}
\usepackage[utf8]{inputenc}

\hypersetup{
colorlinks=true,
linkcolor=blue,
citecolor=blue,
filecolor=magenta,
urlcolor=blue
}
\begin{document}

\title{Thermodynamics and Heat Engine Behavior of Phantom BTZ Black Holes in Noncommutative Geometry }
\author{B. Hamil \orcidlink{0000-0002-7043-6104} \thanks{%
hamilbilel@gmail.com} \\
Laboratoire de Physique Math\'{e}matique et Subatomique,\\
Facult\'{e} des Sciences Exactes, Universit\'{e} Constantine 1, Constantine,
Algeria. \and B. C. L\"{u}tf\"{u}o\u{g}lu 
\orcidlink{Orcid ID :
0000-0001-6467-5005} \thanks{%
bekir.lutfuoglu@uhk.cz (Corresponding author)} \\
Department of Physics, Faculty of Science, University of Hradec Kralove, \\
Rokitanskeho 62/26, Hradec Kralove, 500 03, Czech Republic. }
\date{\today }
\maketitle

\begin{abstract}
This study explores the thermodynamic and geometric properties of phantom BTZ black holes within a noncommutative spacetime framework, where noncommutativity is implemented through Lorentzian smearing of mass and charge distributions. The resulting metric exhibits significant modifications in curvature and horizon structure, particularly in the near-horizon regime. We perform a comparative thermodynamic analysis between phantom and Maxwell field cases, calculating quantities such as Hawking temperature, entropy, heat capacity, and Gibbs free energy. Our findings reveal that noncommutative corrections strongly affect phase transitions and stability conditions. Furthermore, we model the black hole as a heat engine and compute its efficiency, showing how noncommutative effects enhance or suppress energy extraction. This work underscores the interplay between spacetime fuzziness and exotic field dynamics in lower-dimensional gravity, offering new insights into quantum-modified black hole thermodynamics.
\end{abstract}

\\


\section{Introduction}

Black holes, once seen as purely theoretical solutions to Einstein's field equations, are now recognized as critical arenas for probing the intersection of gravity, quantum mechanics, and thermodynamics. A pivotal shift occurred when Bekenstein and Hawking associated entropy with event horizon area and temperature with quantum radiation~\cite{Hawking1972,Bekenstein1973,Bardeen1973,Hawking1975}, inaugurating black hole thermodynamics as a cornerstone of quantum gravity. In $(2+1)$ dimensions, the Bañados–Teitelboim–Zanelli (BTZ) black hole~\cite{Banados1992,Banados1993,Carlip1995} provides an analytically tractable model that retains many essential features of higher-dimensional black holes, such as horizons, Hawking radiation, and thermodynamic behavior~\cite{Carlip1998,Medved2002,Vagenas2002,Zhao2006,Jiang2007,Li2008,Gaete2021}. This has rendered the BTZ solution an effective laboratory for exploring lower-dimensional gravitational dynamics.

At Planckian scales, classical descriptions of spacetime cease to be adequate. Various quantum gravity frameworks—such as string theory and loop quantum gravity—propose a minimal length scale, leading to modifications of the Heisenberg uncertainty principle. Generalized, Extended, and Extended Generalized Uncertainty Principles (GUP, EUP, EGUP) incorporate these quantum corrections and have been fruitfully applied to BTZ thermodynamics~\cite{Zhao2009,Iorio2020,Hamil20221,Hamil20222,Hamil20231}. An alternative route to incorporating short-distance effects lies in noncommutative (NC) geometry~\cite{Seiberg1999}, wherein spacetime coordinates obey nontrivial commutation relations. This approach introduces a natural cutoff and regularizes classical singularities. In black hole physics, NC effects are typically modeled via smeared energy-momentum distributions, yielding modified metrics and altered thermodynamic profiles~\cite{Nicolini2005,Nicolini2006,Rizzo2006,Gingrich2010,Ghosh2018,Anacleto2020,Ovgun2020,Hamil}. In particular, NC corrections have been shown to deform the metric structure~\cite{Dolan2007,Chang2009,Maceda2013}, alter geodesics~\cite{Rahaman2013}, affect scattering amplitudes~\cite{Sadeghi2016}, and reshape entropy-area relations and phase transitions~\cite{Anacleto2018,Gecim2020,Anacleto2021,Anacleto2022,Juric2023}. Additionally, they influence topological and quantum effects such as the Aharonov–Bohm phenomenon~\cite{Anacleto2015}, gauge/gravity correspondence~\cite{Nasseri2005}, and causal structure~\cite{Nozari2008}. The Lorentzian-type smearing of matter and charge, in particular, yields improved regularity and refined thermodynamic behavior~\cite{Liang2012,Ankur2021}. These distributions are rooted in the notion of spatial fuzziness and local anisotropy—concepts that have long been central to models of self-gravitating systems~\cite{Herrera1979}.

Another compelling development in black hole physics involves phantom fields—exotic matter that violates the null energy condition and exhibits negative energy densities~\cite{Caldwell2002,Bronnikov2006}. Their presence can lead to nontrivial features such as horizon shifting, modified curvature profiles, and unconventional thermodynamic behavior~\cite{Clement2009,Rodrigues2012,Gibbons1996,Gyulchev2013}. While concerns regarding quantum instability exist~\cite{Nojiri2003,Piazza2004}, various studies have proposed stable, well-defined phantom configurations~\cite{Hannestad2006,Obs1,Obs2}. When coupled to BTZ spacetimes, phantom energy has been shown to influence entropy scaling, specific heat, and criticality~\cite{Panah2024,Bronnikov2012,Jardim2012,Quevdeo2016}, broadening our understanding of black holes in lower dimensions.

Parallel to these developments, black holes have been reimagined as thermodynamic heat engines~\cite{Johnson2014,Johnson2016,Hennigar2017,Mo2017,Liu2017,Panah2018,Zhang2018,Hendi2018,Mo2018,Chakraborty2019,Wei2019,Balart2019,Zhang2019,Panah2020,Rajani2020,Ghaffarnejad2020,Promsiri2021,Roy2021,Panah2022,Cao2022,DiMarco2023,Nag2023,Javed2024,Kruglov2024,Panah2025,Mondal2025,Fatima2025}. Introduced by Johnson, this framework extends the standard thermodynamic analogy by treating the cosmological constant as a dynamical pressure and its conjugate as volume, enabling the construction of pressure–volume ($P$–$V$) cycles analogous to those in classical heat engines. In this picture, black holes function as working substances that convert heat input into mechanical work, and their efficiency can be computed for specific thermodynamic trajectories (e.g., rectangular cycles in the $P$–$V$ plane). The BTZ black hole, due to its lower-dimensional simplicity and analytic tractability, has emerged as a compelling platform for studying such heat engines~\cite{Mo2017,Balart2019}. Extensions have examined the influence of diverse physical phenomena on engine performance, including the effects of dark energy~\cite{Liu2017}, gravity’s rainbow~\cite{Panah2018}, massive gravity~\cite{Hendi2018}, and quantum corrections \cite{Fatima2025}. Modifications of the thermodynamic variables have been shown to affect the critical behavior, phase structure, and energy extraction efficiency. Moreover, nonstandard black hole configurations—such as Bardeen~\cite{Rajani2020,Nag2023}, hairy~\cite{Ghaffarnejad2020}, accelerating~\cite{Zhang2019}, and cavity-confined geometries~\cite{Cao2022}—have broadened the applicability of this paradigm. Recent works have also focused on benchmarking techniques~\cite{Chakraborty2019}, universal features in critical exponents~\cite{DiMarco2023}, and thermodynamic geometry approaches to characterize engine performance and phase transitions~\cite{Promsiri2021,Panah2022}. Even more exotic setups—including acoustic analogues~\cite{Mondal2025}, ModMax-AdS black holes~\cite{Panah2025}, and magnetically charged configurations~\cite{Kruglov2024}—have confirmed the versatility of the heat engine formulation across a wide spectrum of gravitational systems. These efforts collectively demonstrate that the black hole heat engine is not merely a theoretical construct, but a powerful tool for probing the interplay between gravity, thermodynamics, and quantum corrections.

Motivated by these intersections, this study investigates BTZ black holes within noncommutative geometry in the presence of both phantom and Maxwell fields. Using Lorentzian-smeared energy-momentum tensors, we derive a modified metric and analyze its geometric and thermodynamic properties, including Hawking temperature, entropy, heat capacity, and Gibbs free energy. These quantities are examined to uncover phase transitions and assess the stability of the solutions. We further evaluate the black hole's performance as a thermodynamic heat engine, quantifying the effects of noncommutativity and exotic matter content on energy conversion efficiency—an aspect not previously explored for these configurations. In fact, this work synthesizes and extends two prior directions in the literature: while Ref.~\cite{Ankur2021} studied dynamic noncommutative BTZ black holes using both Gaussian and Lorentzian smearing functions—without coupling to exotic matter—and Ref.~\cite{Panah2024} analyzed BTZ black holes in the presence of both phantom and Maxwell fields—without invoking noncommutative geometry—our approach bridges these frameworks. By incorporating Lorentzian-type noncommutative corrections into BTZ black holes that include both phantom and Maxwell fields, and by conducting a comparative analysis of their heat engine behavior, we provide a unified platform for exploring the interplay between quantum spacetime structure and exotic matter in $(2+1)$-dimensional gravity.

This paper is structured as follows: Section~\ref{sec2} presents the construction of the noncommutative phantom BTZ metric. Section~\ref{sec3} details its thermodynamic properties and stability. Section~\ref{sec4} analyzes the black hole as a heat engine. Conclusions and outlook are provided in Section~\ref{sec5}.



\section{Phantom BTZ black holes in noncommutative space} \label{sec2}

In a noncommutative spacetime, non-commutativity is introduced in the Cartesian coordinate system through the following commutation relations:
\begin{equation}
\left[ x^{\mu },x^{\nu }\right] =i\Theta ^{\mu \nu }.
\end{equation}%
Here, $\Theta ^{\mu \nu }$ is a constant, antisymmetric tensor with dimensions of (length)$^{2}$. The effects of non-commutativity gradually diminish and vanish as the noncommutative parameter $\Theta$ approaches zero. 

\begin{figure}[H]
\begin{minipage}[t]{0.5\textwidth}
        \centering
        \includegraphics[width=\textwidth]{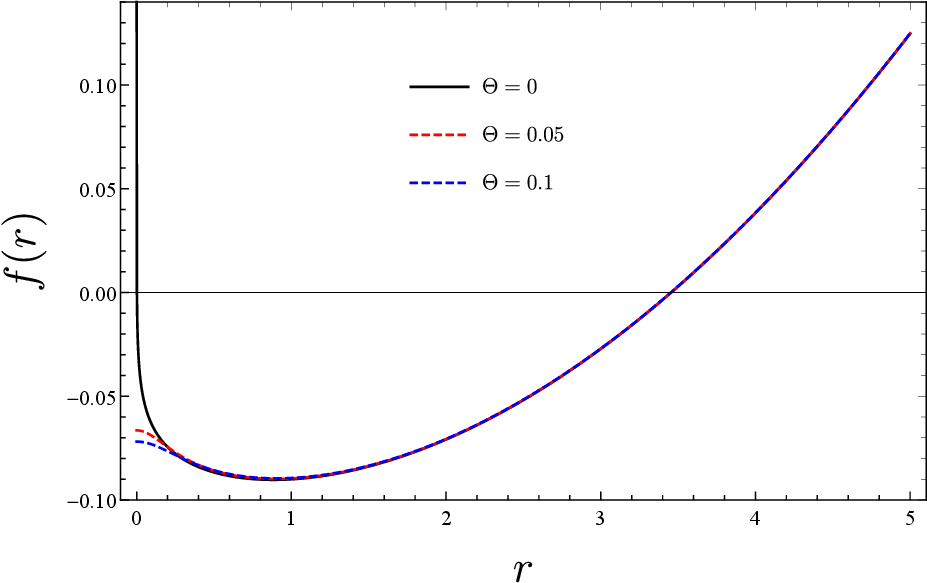}
                \subcaption{$\eta=1$}
        \label{fig:met1}
\end{minipage}
\begin{minipage}[t]{0.5\textwidth}
        \centering
        \includegraphics[width=\textwidth]{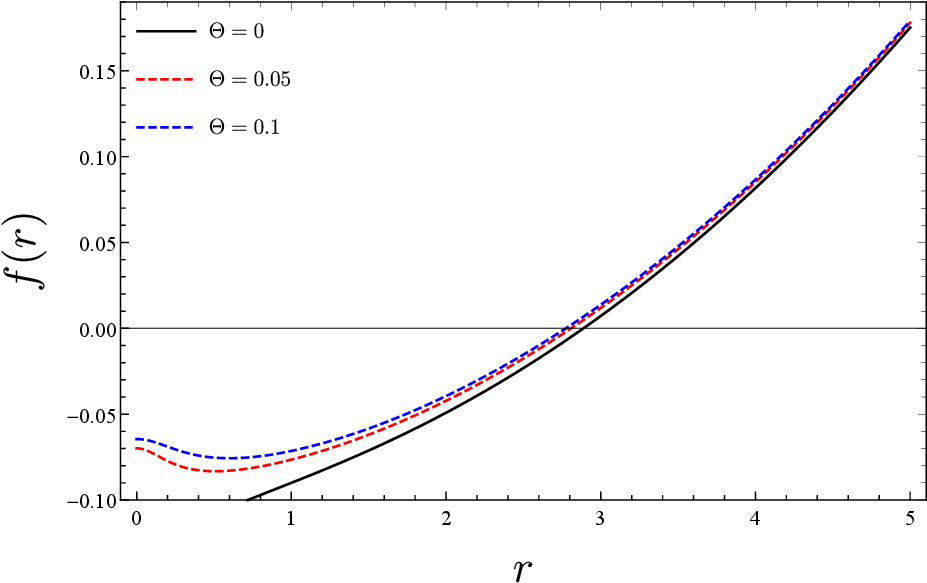}
                \subcaption{$\eta=-1$}
        \label{fig:met2}
\end{minipage}
\caption{Variation of the metric function with respect to the event horizon radius $r_H$ for $\Lambda=-10^{-2}$,  with $r_0=1$, $M=0.1$ and $q=\frac{\pi}{2}$. }
\label{fig:met}
\end{figure}

Non-commutativity of spacetime eliminates point-like sources by spreading an object over space \cite{Seiberg1999}, thereby affecting the propagation of its energy and momentum \cite{Herrera1979}. This spacetime fuzziness results in significant changes to the distributions of a particle’s mass and charge. To account for these modifications, the metric of the noncommutative phantom BTZ black hole is derived by replacing Dirac’s point-like mass and charge distributions with a distribution function of minimal width $\sqrt{\Theta}$. Following \cite{Ankur2021}, we consider the mass and charge distributions that are described by Lorentzian functions in $(2+1)$ dimensions:
\begin{equation}
\rho _{matt}\left( r,\Theta \right) =\frac{M\sqrt{\Theta }}{2\pi \left(
r^{2}+\Theta \right) ^{3/2}},
\end{equation}%
\begin{equation}
\rho _{el}\left( r,\Theta \right) =\frac{q\sqrt{\Theta }}{2\pi \left(
r^{2}+\Theta \right) ^{3/2}},
\end{equation}%
where $M$ and $q$ denote the mass and charge of the black hole, respectively. The action governing phantom BTZ black holes in three-dimensional spacetime is given by \cite{Panah2024}:
\begin{equation}
\mathcal{I}=\frac{1}{16\pi G}\int d^{3}x\sqrt{-g}\left( R+2\Lambda +\eta \mathcal{F}%
\right).  \label{1}
\end{equation}%
Here, $R$ represents the Ricci scalar, $\Lambda$ denotes the cosmological constant, and $\mathcal{F}=F_{\mu \nu }F^{\mu \nu }$ corresponds to the Maxwell invariant. The parameter $\eta$ distinguishes between the Maxwell field $(\eta = 1)$ and the phantom field $(\eta = -1)$. Now, we apply the variational principle from Eq.~(\ref{1}) to derive the Einstein equations:
\begin{equation}
R_{\mu \nu }-\frac{1}{2}g_{\mu \nu }R+\Lambda g_{\mu \nu }=\pi \left( \left.
T_{\mu \nu }\right\vert _{matt}+\eta \left. T_{\mu \nu }\right\vert
_{el}\right) ,  \label{Ein}
\end{equation}
\begin{equation}
\frac{1}{\sqrt{-g}}\partial _{\mu }\left( \sqrt{-g}F^{\mu \nu }\right)
=J^{\nu },  \label{e1}
\end{equation}
where $\left. T_{\mu \nu }\right\vert _{matt}$ is given as in \cite{Liang2012}:
\begin{equation}
\left. T_{\mu }^{\nu }\right\vert _{matt}=diag\left( 
\begin{array}{ccc}
-\rho _{matt} & p_{r} & p_{t}%
\end{array}%
\right) ,
\end{equation}
with the radial and tangential pressures defined as $p_{r}=-\rho_{matt}$ and $p_{t}=-\rho_{matt}-r\partial \rho_{matt}$. The electromagnetic stress-energy tensor is expressed in the standard form:
\begin{equation}
\left. T_{\mu \nu }\right\vert _{el}=-\frac{1}{4\pi }\left( F_{\mu \alpha}g^{\alpha \beta}F_{\beta \nu }-\frac{1}{4}g_{\mu \nu }F_{\sigma \alpha}g^{\alpha \beta}F_{\beta \rho }g^{\rho \sigma}\right).
\end{equation}
The charge distribution is assumed to be static, with the current density $J^{\nu}$ being nonzero only in the time direction:
\begin{equation}
J^{\nu }=\rho_{el}(r,\Theta)\delta_{0}^{\nu }.  \label{e2}
\end{equation}
The nonzero components of the field strength are:
\begin{equation}
F^{r0}=-F^{0r}=E(r,\Theta).  \label{3}
\end{equation}
Substituting Eqs.~(\ref{e2}) and (\ref{3}) into Eq.~(\ref{e1}) and solving the Maxwell equation, the electric field is obtained as:
\begin{equation}
E(r,\Theta) = \frac{q}{2\pi r} \left( 1-\frac{\sqrt{\Theta }}{\sqrt{r^{2}+\Theta }}\right).
\end{equation}
We consider a three-dimensional static spacetime described by the line element:
\begin{equation}
ds^{2}=-f(r)dt^{2}+\frac{1}{f(r)}dr^{2}+r^{2}d\varphi^{2},
\end{equation}
where the metric function $f(r)$ is defined as:
\begin{equation}
f(r) = \Psi(r,\Theta) - \Lambda r^{2}.  \label{mat}
\end{equation}
To determine $f(r)$, we use the time component of Eq.~(\ref{Ein}), yielding:
\begin{equation}
\frac{\partial}{\partial r}\Psi(r,\Theta) + \left[ \frac{M\sqrt{\Theta }}{\left( r^{2}+\Theta \right)^{3/2}}r + \frac{\eta}{4}\frac{q^{2}}{4\pi^{2}r}\left( 1-\frac{\sqrt{\Theta }}{\sqrt{r^{2}+\Theta }}\right)^{2}\right] = 0.
\end{equation}
Integrating this equation, we find:
\begin{equation}
\Psi(r,\Theta) = c_{1} + \frac{\sqrt{\Theta }M}{\pi \sqrt{\Theta + r^{2}}} 
- \frac{\eta q^{2}}{8\pi^{2}} \left[ \log\bigg(\frac{ \sqrt{\Theta }+\sqrt{\Theta +r^{2}}}{r_0}\bigg)
- \frac{1}{4}\log\bigg(\frac{ \Theta +r^{2}}{r_0^2}\bigg) \right]. \label{psifunc}
\end{equation}
The integration constant $c_1$ is determined by demanding that the metric reduces to the standard BTZ black hole form in the commutative limit $\Theta \to 0$, thereby identifying $c_1$ with the black hole mass parameter $-M$. In addition, $r_0$ appears as an integration constant that serves as a reference length scale, ensuring that the arguments of the logarithmic terms remain dimensionless. Substituting Eq.~(\ref{psifunc}) into Eq.~(\ref{mat}), the metric function for the phantom BTZ black hole in noncommutative geometry is:
\begin{equation}
f(r) = -M + \frac{\sqrt{\Theta }M}{\pi \sqrt{\Theta + r^{2}}} 
- \frac{\eta q^{2}}{8\pi^{2}} \left[ \log\bigg(\frac{ \sqrt{\Theta }+\sqrt{\Theta +r^{2}}}{r_0}\bigg) 
- \frac{1}{4}\log\bigg(\frac{ \Theta +r^{2}}{r_0^2}\bigg) \right] - \Lambda r^{2}. \label{metricfunc}
\end{equation}
Figure \ref{fig:met} presents the metric functions for both the standard (panel (a)) and phantom (panel (b)) BTZ black holes in noncommutative spacetime. 

\noindent In both cases, noncommutativity corrections primarily affect the small-$r$ region while leaving the asymptotic structure unchanged. For the standard BTZ black hole, increasing $\Theta$ slightly modifies the near-horizon behavior, suggesting a regularization effect that may influence thermodynamic properties such as temperature and entropy. In contrast, the phantom BTZ case exhibits a more pronounced deviation, particularly in the slope of the metric function near $r = 0$, indicating stronger alterations in the causal structure. The presence of exotic matter in the phantom scenario, combined with NC effects, could lead to significant changes in horizon structure and stability. In particular, a greater noncommutativity parameter leads to a reduced horizon radius.  In both cases, the close agreement of the curves at large $r$ confirms that noncommutativity introduces only localized modifications, affecting the short-distance properties of the black hole without altering its asymptotic behavior. These findings highlight the role of noncommutativity in modifying the inner geometry of BTZ black holes, with potential implications for their thermodynamics and singularity structure.

Expanding Eq. \eqref{metricfunc} to the first order in $\Theta$, the metric function becomes:
\begin{equation}
f(r) = -M\left( 1-\frac{\sqrt{\Theta }}{\pi r}\right) 
- \frac{\eta q^{2}}{8\pi^{2}} \left[ \frac{1}{2} \log\left( \frac{r}{r_{0}}\right) 
+ \frac{\sqrt{\Theta }}{r} - \frac{\Theta}{4r^{2}}\right] - \Lambda r^{2}. \label{metric}
\end{equation}
To analyze the geometrical structure of these solutions, we begin by examining the presence of essential singularities through the computation of the Ricci and Kretschmann scalars. These scalars are derived and expressed as follows:
\begin{equation}
R = 6\Lambda + \frac{\eta q^{2}\left( r^{2} - \Theta \right)}{16\pi^{2}r^{4}},
\end{equation}
\small
\begin{eqnarray}
K = 12\Lambda^{2} + \frac{3q^{4}}{256\pi^{4}r^{4}} + \frac{\eta \Lambda q^{2}}{4\pi^{2}r^{2}} 
+ \sqrt{\Theta} \frac{\eta q^{2}}{2\pi^{3}r^{5}} \left( M - \frac{\eta q^{2}}{8\pi} \right) 
+ \frac{\Theta}{\pi^{2}r^{2}} \left( \frac{6M^{2}}{r^{4}} - \frac{\eta \Lambda q^{2}}{4r^{2}} 
- \frac{3M\eta q^{2}}{2\pi r^{4}} + \frac{17q^{4}}{128\pi^{2}r^{4}} \right).
\end{eqnarray} \normalsize
Both the Ricci scalar $R$ and the Kretschmann scalar $K$ exhibit divergences at the origin, indicating the presence of a curvature singularity:
\begin{equation}
\lim_{r\rightarrow 0} R = \infty,
\end{equation}
\begin{equation}
\lim_{r\rightarrow 0} K = \infty.
\end{equation}
Thus, a curvature singularity exists at $r = 0$. For large radial distances $(r \to \infty)$, the scalars simplify to:
\begin{equation}
\lim_{r\rightarrow \infty} R = 6\Lambda,
\end{equation}
\begin{equation}
\lim_{r\rightarrow \infty} K = 12\Lambda^{2}.
\end{equation}
This behavior indicates that the spacetime becomes independent of the parameters $(\eta, \Theta)$ at large distances and is asymptotically anti-de Sitter (AdS).


\section{Thermodynamic Analysis} \label{sec3}
In this section, we investigate the thermodynamic properties of both ordinary ($\eta=1$) and phantom ($\eta=-1$) BTZ black holes within the framework of noncommutative geometry. To achieve this, we express the black hole mass $M$ in terms of the event horizon radius $r_H$, the cosmological constant $\Lambda$, the charge $q$, and the noncommutative parameter $\Theta$, by imposing the condition $f(r_H)=0$. The mass is then given by the following expression:
\begin{eqnarray}
M &=&-\frac{\eta q^{2}}{16\pi ^{2}}\log \Big(\frac{r_{H}}{r_{0}}\Big)-\Lambda r_{H}^{2}-\frac{\sqrt{\Theta }}{\pi r_{H}}\left( \Lambda r_{H}^{2}+\frac{\eta q^{2}}{8\pi }+\frac{\eta q^{2}}{16\pi ^{2}}\log \Big(\frac{r_{H}}{r_{0}}\Big)\right)   \notag \\ &+&\frac{\Theta }{\pi ^{2}r_{H}^{2}}\left( \frac{(\pi -4)\eta q^{2}}{32\pi }-\Lambda r_{H}^{2}-\frac{\eta q^{2}}{16\pi ^{2}}\log \Big(\frac{r_{H}}{r_{0}}\Big)\right) .  \label{mass}
\end{eqnarray}
This expression incorporates first-order corrections in the noncommutative parameter $\Theta$, which introduces modifications to the classical behavior of the black hole mass.  

\begin{figure}[H]
\begin{minipage}[t]{0.5\textwidth}
        \centering
        \includegraphics[width=\textwidth]{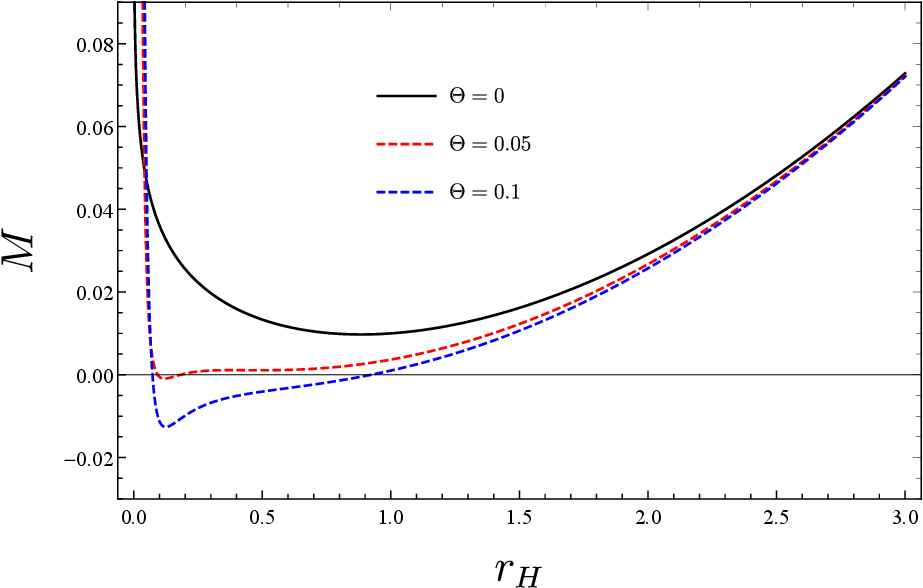}
                \subcaption{$\Lambda=-10^{-2}$}
        \label{fig:mass11}
\end{minipage}
\begin{minipage}[t]{0.5\textwidth}
        \centering
        \includegraphics[width=\textwidth]{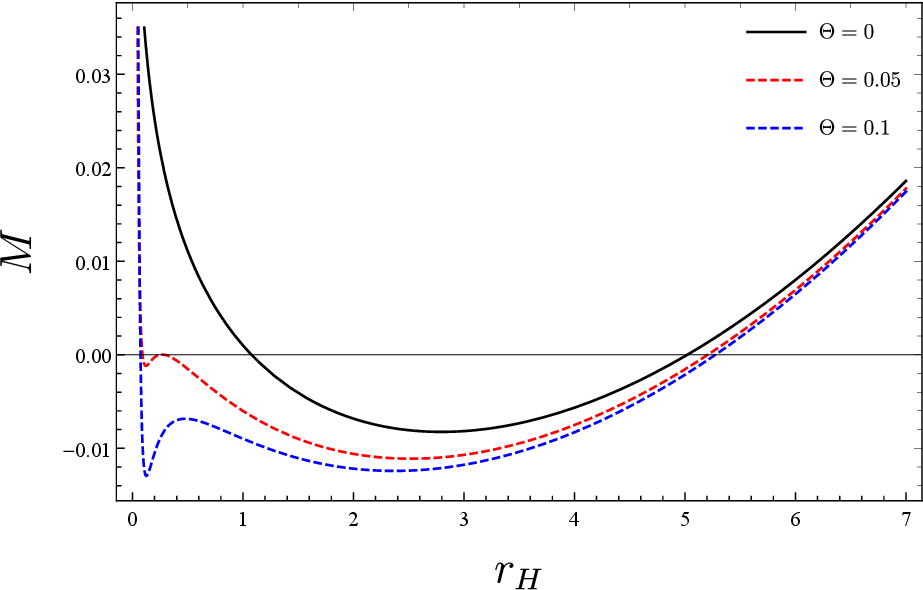}
                \subcaption{$\Lambda=-10^{-3}$}
        \label{fig:mass12}
\end{minipage}
\caption{BTZ black hole mass $M$ as a function of the event horizon radius $r_H$ for $\eta=1$, with $r_0=1$ and $q=\frac{\pi}{2}$. Panel (a) corresponds to $\Lambda = -10^{-2}$, and panel (b) corresponds to $\Lambda = -10^{-3}$.}
\label{fig:mass1}
\end{figure}

\noindent Figure \ref{fig:mass1} presents the BTZ black hole mass $M$ as a function of the event horizon radius $r_H$ in noncommutative spacetime for $\Lambda = -10^{-2}$ (panel (a)) and $\Lambda = -10^{-3}$ (panel (b)). In both cases, noncommutativity, parameterized by $\Theta$, introduces deviations in the small-$r_H$ regime, with larger $\Theta$ leading to more pronounced corrections. These effects are more noticeable for smaller $\Lambda$, as seen in panel (b), where the slower variation of $M$ with $r_H$ enhances the visibility of NC corrections. At large $r_H$, all curves converge, confirming that noncommutativity primarily affects the near-horizon structure while preserving the asymptotic behavior. The observed modifications suggest potential implications for the thermodynamic properties and stability of the black hole, particularly in the small-$r_H$ region.

\begin{figure}[H]
\begin{minipage}[t]{0.5\textwidth}
        \centering
        \includegraphics[width=\textwidth]{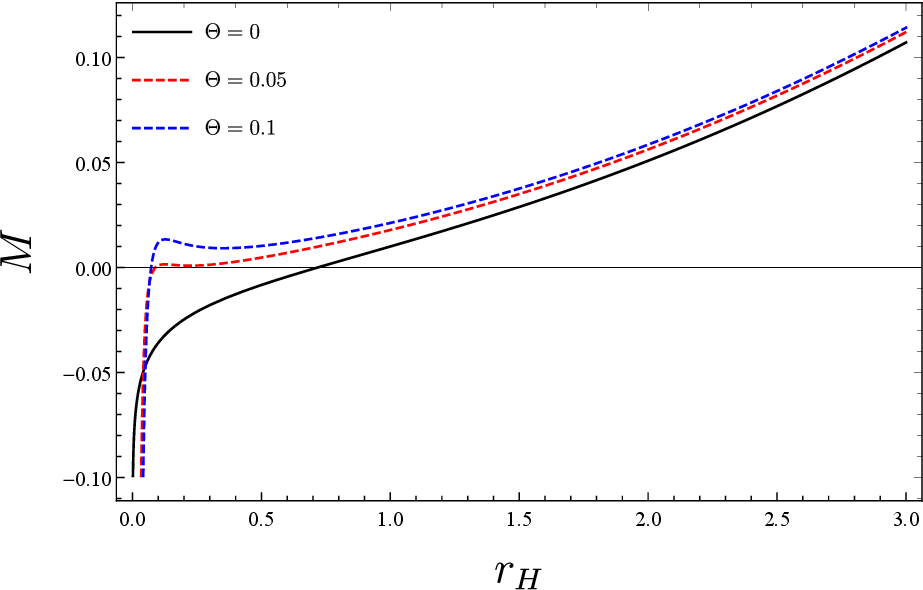}
                \subcaption{$\Lambda=-10^{-2}$}
        \label{fig:mass21}
\end{minipage}
\begin{minipage}[t]{0.5\textwidth}
        \centering
        \includegraphics[width=\textwidth]{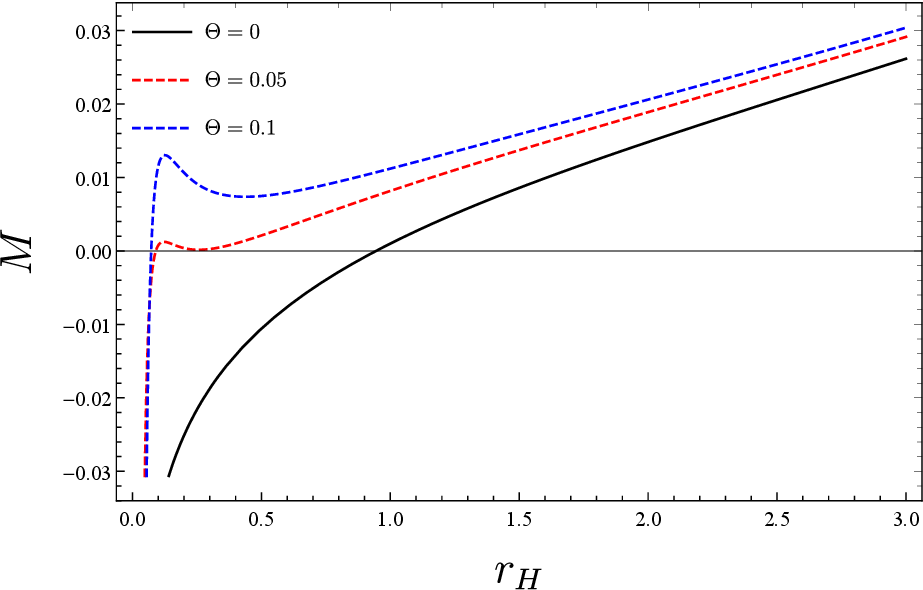}
                \subcaption{$\Lambda=-10^{-3}$}
        \label{fig:mass22}
\end{minipage}
\caption{BTZ black hole mass $M$ as a function of the event horizon radius $r_H$ for $\eta=-1$, with $r_0=1$ and $q=\frac{\pi}{2}$. Panel (a) corresponds to $\Lambda = -10^{-2}$, and panel (b) corresponds to $\Lambda = -10^{-3}$.}
\label{fig:mass2}
\end{figure}

Figure~\ref{fig:mass2} depicts the dependence of the phantom BTZ black hole mass on the event horizon radius for different values of the noncommutative deformation parameter. The phantom nature implies that the mass function can take negative values, contrasting with the standard BTZ case. The undeformed solution $\Theta = 0$ serves as a baseline, while noncommutative effects $\Theta > 0$ introduce deviations that are more pronounced for small $r_H$. These corrections suggest that noncommutativity alters the near-horizon geometry, potentially modifying the causal structure. At larger $r_H$, the mass asymptotically approaches the undeformed behavior, indicating that noncommutative effects are localized near the event horizon. A smaller absolute value of the cosmological constant (panel b) results in a less steep mass function, reflecting the suppression of AdS curvature effects. The results highlight that in the phantom BTZ scenario, noncommutative corrections become significant in the deep infrared regime, possibly influencing the stability and thermodynamic properties of the black hole.

\begin{figure}[H]
\begin{minipage}[t]{0.5\textwidth}
        \centering
        \includegraphics[width=\textwidth]{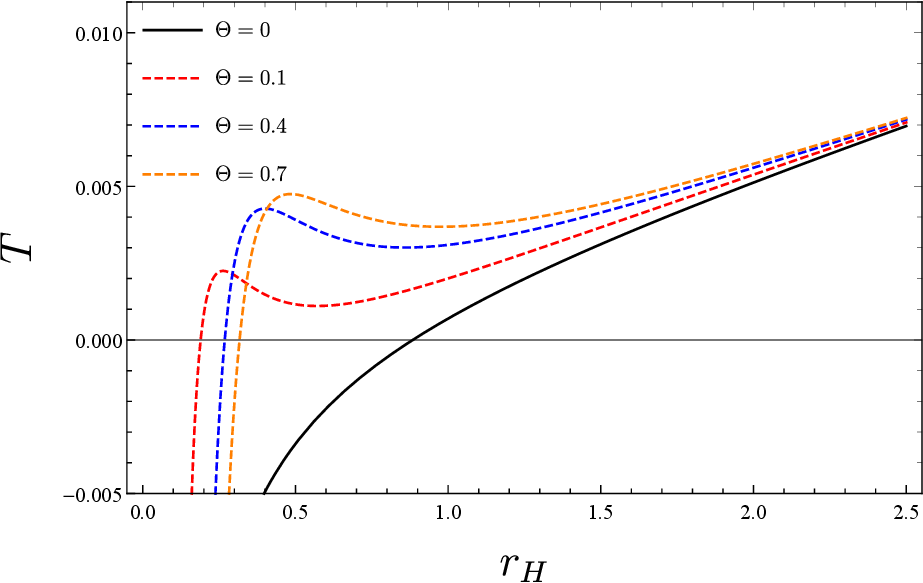}
                \subcaption{$\Lambda=-10^{-2}$}
        \label{fig:tem1}
\end{minipage}
\begin{minipage}[t]{0.5\textwidth}
        \centering
        \includegraphics[width=\textwidth]{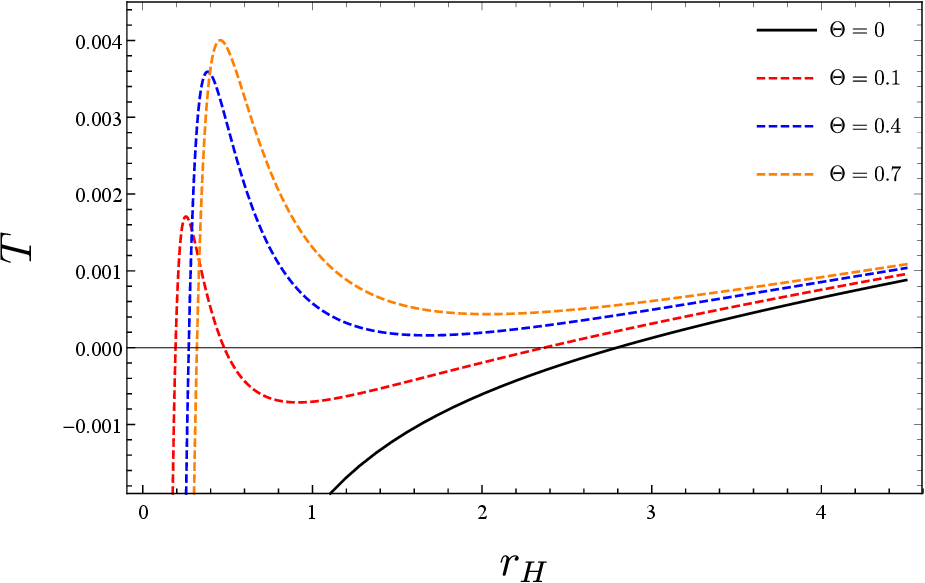}
                \subcaption{$\Lambda=-10^{-3}$}
        \label{fig:tem2}
\end{minipage}
\caption{Hawking temperature $T$ as a function of the event horizon radius $r_H$ for $\eta=1$, with $r_0=1$ and $q=\frac{\pi}{2}$. Panel (a) corresponds to $\Lambda = -10^{-2}$, and panel (b) corresponds to $\Lambda = -10^{-3}$.}
\label{fig:Htem1}
\end{figure}

\begin{figure}[H]
\begin{minipage}[t]{0.5\textwidth}
        \centering
        \includegraphics[width=\textwidth]{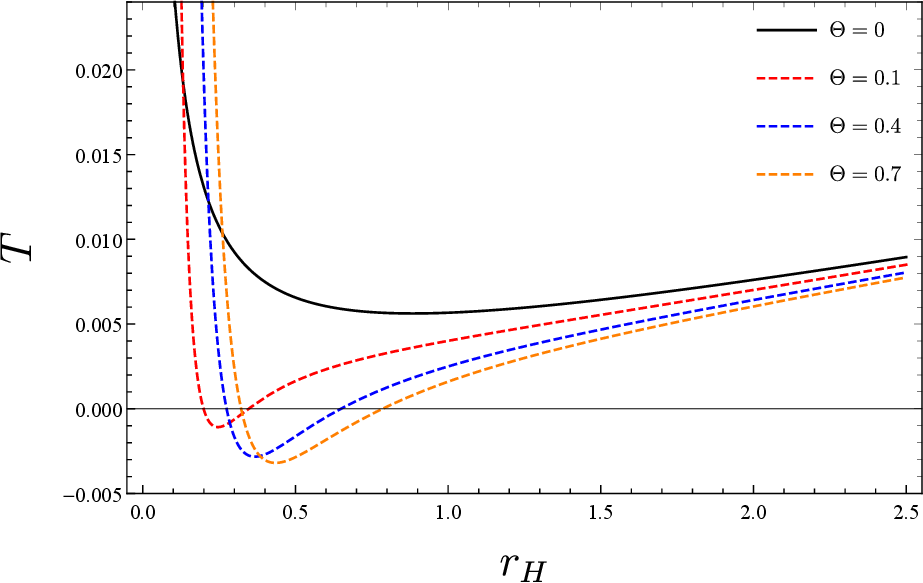}
                \subcaption{$\Lambda=-10^{-2}$}
        \label{fig:tem3}
\end{minipage}
\begin{minipage}[t]{0.5\textwidth}
        \centering
        \includegraphics[width=\textwidth]{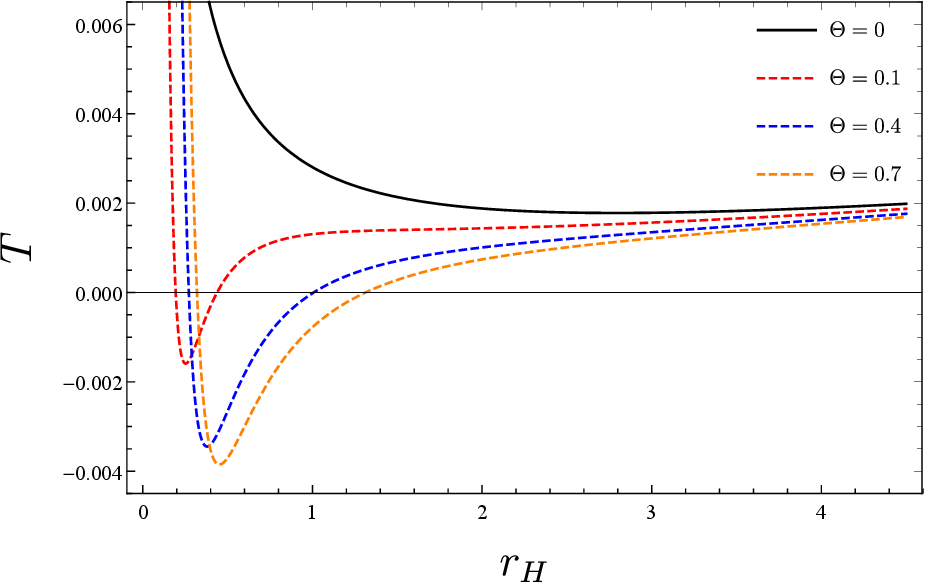}
                \subcaption{$\Lambda=-10^{-3}$}
        \label{fig:tem4}
\end{minipage}
\caption{Hawking temperature $T$ as a function of the event horizon radius $r_H$ for $\eta=-1$, with $r_0=1$ and $q=\frac{\pi}{2}$. Panel (a) corresponds to $\Lambda = -10^{-2}$, and panel (b) corresponds to $\Lambda = -10^{-3}$.}
\label{fig:Htem2}
\end{figure}

Based on the given metric, we now calculate the Hawking temperature using the following relation:
\begin{equation}
T=\frac{1}{4\pi }\left. \frac{d}{dr}f\left( r\right) \right\vert _{r=r_{H}}. \label{17}
\end{equation}%
After evaluating the derivative at the horizon radius, the mass expression from Eq.~\eqref{mass} is substituted to obtain the Hawking temperature:
\begin{eqnarray}
T &=&-\frac{\eta q^{2}}{32\pi ^{3}r_{H}}-\frac{\Lambda r_{H}}{\pi }+\frac{%
\sqrt{\Theta }}{2\pi ^{2}r_{H}^{2}}\left( \frac{\eta q^{2}}{8\pi }+\Lambda
r_{H}^{2}+\frac{\eta q^{2}}{16\pi ^{2}}\log \Big(\frac{r_{H}}{r_{0}}\Big)\right)  
\notag \\
&+&\frac{\Theta }{2\pi ^{3}r_{H}^{3}}\left( \Lambda r_{H}^{2}-\frac{(\pi
-2)\eta q^{2}}{16\pi }+\frac{\eta q^{2}}{16\pi ^{2}}\log \Big(\frac{r_{H}}{r_{0}}\Big)\right) . \label{etem}
\end{eqnarray}%
To qualitatively analyze the effect of noncommutative corrections, we plot the Hawking temperature as a function of the event horizon radius. The resulting plots are shown in Figure \ref{fig:Htem1} and Figure \ref{fig:Htem2}.

For the case $\eta = 1$ (normal case), in the commutative space limit (i.e., $\Theta = 0$), the Hawking temperature $T$ increases monotonically with the event horizon radius $r_H$, consistent with classical thermodynamics of BTZ black holes. However, in the noncommutative case (i.e., $\Theta \neq 0$), this monotonic increase is modified. For relatively smaller BTZ black holes, noncommutative corrections cause the temperature $T$ to exhibit a non-monotonic behavior, including the appearance of a dip or a plateau at small $r_H$ before eventually increasing at larger $r_H$. This deviation from monotonicity becomes more pronounced as the noncommutative parameter $\Theta$ increases, reflecting the significant influence of noncommutativity on the thermodynamics of small black holes. Physically, this behavior suggests that noncommutativity introduces a stabilizing effect, potentially preventing divergences in $T$ for small horizon radii while also altering the thermodynamic profile of the black hole. For larger black holes (larger $r_H$), the influence of noncommutativity diminishes, and $T$ approaches the classical behavior seen in the commutative limit.
For the case $\eta = -1$ (phantom case), the thermodynamic behavior of the Hawking temperature $T$ is significantly influenced by both phantom energy and noncommutative corrections. In the commutative space limit (i.e., $\Theta = 0$), the temperature $T$ does not exhibit a clear monotonic increase with the event horizon radius $r_H$. Instead, $T$ shows non-monotonic behavior, potentially involving a dip or plateau at small $r_H$, reflecting the influence of phantom energy on the thermodynamic properties of the black hole. When noncommutative corrections are introduced (i.e., $\Theta \neq 0$), this non-monotonic behavior is further modified. For small $r_H$, $T$ can become negative, indicating thermodynamic instability induced by the combined effects of phantom energy and noncommutativity. This negative temperature regime becomes more pronounced as $\Theta$ increases, highlighting the destabilizing role of noncommutative corrections for small BTZ black holes. For larger $r_H$, however, $T$ recovers to positive values and asymptotically approaches classical behavior as the influence of noncommutativity diminishes. These results illustrate how noncommutativity amplifies the exotic thermodynamic features of phantom energy at small scales while stabilizing the system at larger scales.

Next, we apply the first law of black hole thermodynamics, 
\begin{equation}
 dS=\frac{1}{T}%
\frac{\partial M}{\partial r_{H}}dr_{H},   
\end{equation}
to derive the Bekenstein entropy.
Utilizing Eqs. (\ref{mass}) and (\ref{etem}), the entropy is expressed as: 
\begin{equation}
S=\pi r_{H}+2\sqrt{\Theta }\log \Big(\frac{r_{H}}{r_{0}}\Big)-\frac{2\Theta }{\pi r_{H}}. \label{ent}
\end{equation}
This result demonstrates that the Bekenstein entropy is modified by the noncommutative parameter, while the entropy remains identical for both phantom and Maxwell black holes. Building on this, we turn our attention to another key thermodynamic quantity---the heat capacity---which provides critical insights into the stability and phase transitions of the black hole system. The heat capacity $C$ is defined as%
\begin{equation}
C=T\left( \frac{\partial S}{\partial T}\right) _{q}.
\end{equation}
Using Eqs.~(\ref{etem}) and (\ref{ent}), the heat capacity can be explicitly calculated as:
\begin{eqnarray}
C &=&2\pi r_{H} \frac{\left( \Lambda r_{H}^{2}+\frac{\eta q^{2}}{32\pi ^{2}}\right) }{\left( \Lambda r_{H}^{2}-\frac{\eta q^{2}}{32\pi ^{2}}\right) } \notag \\
&-&\frac{\sqrt{\Theta }}{\left( \Lambda r_{H}^{2}-\frac{\eta q^{2}}{32\pi ^{2}}\right) ^{2}}\Bigg[\frac{q^{4}}{256\pi ^{3}}+\frac{3\eta q^{2}}{16\pi ^{2}} \left( \frac{\eta q^{2}}{96\pi ^{2}}+\Lambda r_{H}^{2}\right) \log \Big(\frac{r_{H}}{r_{0}}\Big)+\frac{3(4\pi -1)}{32\pi ^{2}}\eta \Lambda q^{2}r_{H}^{2}-\Lambda ^{2}r_{H}^{4}\Bigg]  \notag \\
&+&\frac{3}{\left( 32\pi ^{2}\right) ^{2}\pi r\left( \Lambda r_{H}^{2}-\frac{\eta q^{2}}{32\pi ^{2}}\right) ^{3}}\Bigg[q^{4}\left( \frac{\eta q^{2}}{96\pi ^{2}}+\Lambda r_{H}^{2}\right) \log ^{2}\Big(\frac{r_{H}}{r_{0}}\Big)  \notag \\
&-&32\pi ^{2}\eta q^{2}\left( -\frac{q^{4}}{768\pi ^{3}}+\frac{(1-12\pi )}{96\pi ^{2}}\eta \Lambda q^{2}r_{H}^{2}+\Lambda ^{2}r_{H}^{4}\right) \log \Big(\frac{r_{H}}{r_{0}}\Big)  \notag \\
&+&\frac{256\pi ^{4}}{3}\left( \frac{3\eta q^{6}}{8192\pi ^{4}}+\frac{(11\pi-2)}{256\pi ^{3}}\Lambda q^{4}r_{H}^{2}+\frac{\left( 5-24\pi +8\pi^{2}\right) }{32\pi ^{2}}\eta \Lambda ^{2}q^{2}r_{H}^{4}-\Lambda^{3}r_{H}^{6}\right) \Bigg].  \label{heat5}
\end{eqnarray}
The heat capacity plays a pivotal role in determining the local thermodynamic stability of the black hole. When $C > 0$, the black hole is locally stable, whereas $C < 0$ indicates thermodynamic instability. Additionally, the heat capacity diverges at points where its denominator vanishes, signaling phase transitions. Specifically, the heat capacity becomes infinite at:
\begin{equation}
    r_H = \sqrt{\frac{\eta q^2}{32\Lambda \pi^2}},
\end{equation}
indicating the presence of a critical radius where a phase transition occurs. 

\begin{figure}[H]
\begin{minipage}[t]{0.5\textwidth}
        \centering
        \includegraphics[width=\textwidth]{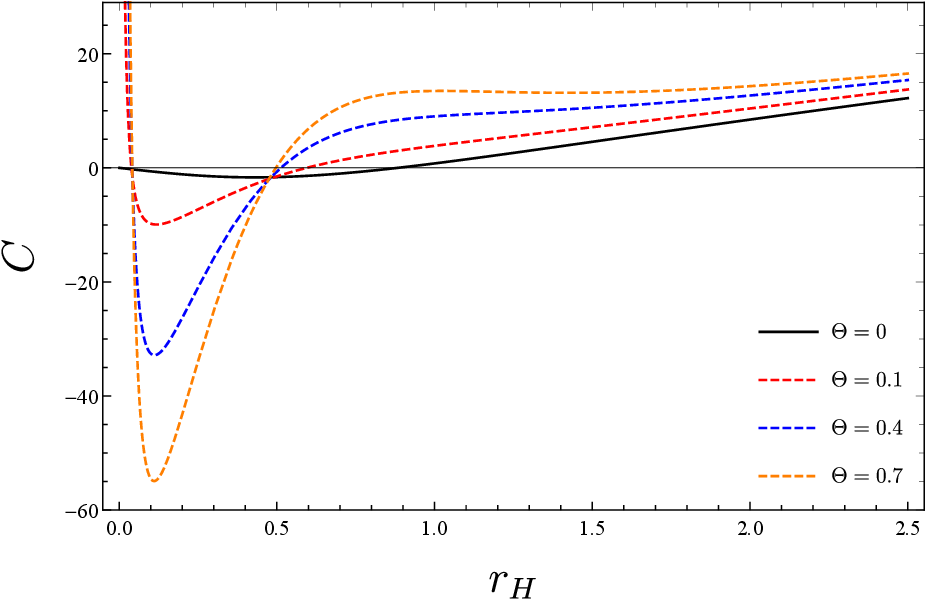}
                \subcaption{$\Lambda=-10^{-2}$}
        \label{fig:heat1}
\end{minipage}
\begin{minipage}[t]{0.5\textwidth}
        \centering
        \includegraphics[width=\textwidth]{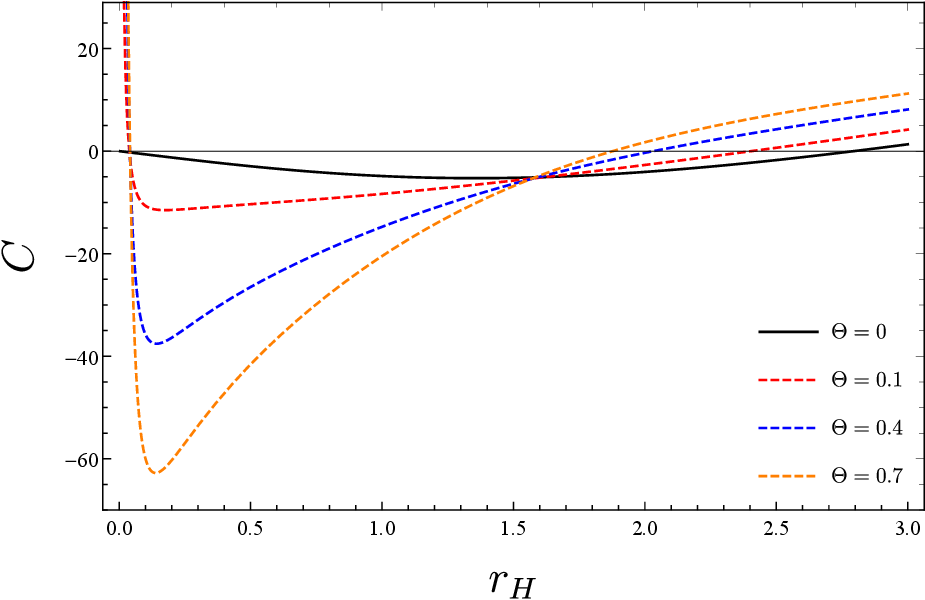}
                \subcaption{$\Lambda=-10^{-3}$}
        \label{fig:heat2}
\end{minipage}
\caption{Heat capacity $C$ as a function of the event horizon radius $r_H$ for $\eta = 1$, with $r_0=1$ and $q=\frac{\pi}{2}$.}
\label{fig:Hheat1}
\end{figure}

\begin{figure}[H]
\begin{minipage}[t]{0.5\textwidth}
        \centering
        \includegraphics[width=\textwidth]{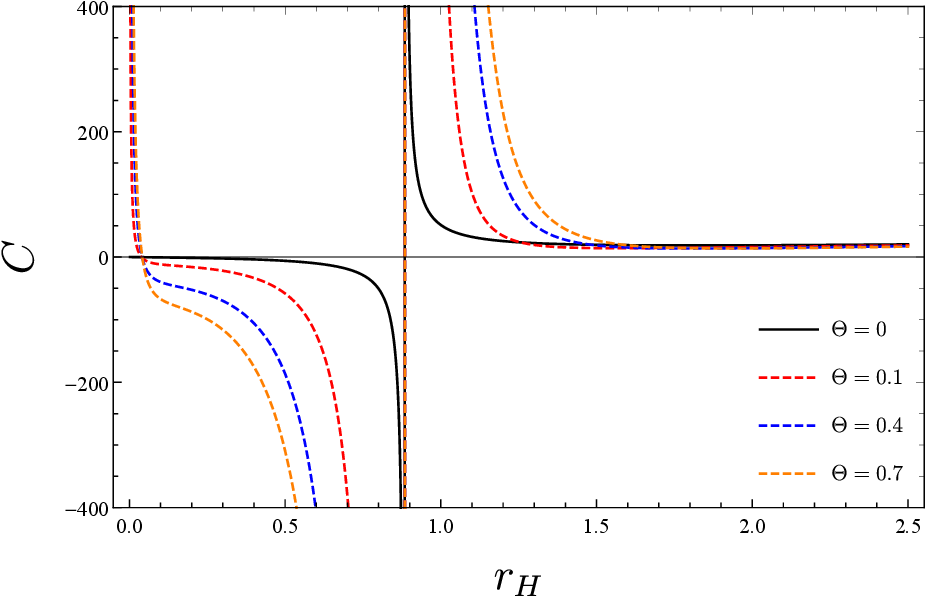}
                \subcaption{$\Lambda=-10^{-2}$}
        \label{fig:heat3}
\end{minipage}
\begin{minipage}[t]{0.5\textwidth}
        \centering
        \includegraphics[width=\textwidth]{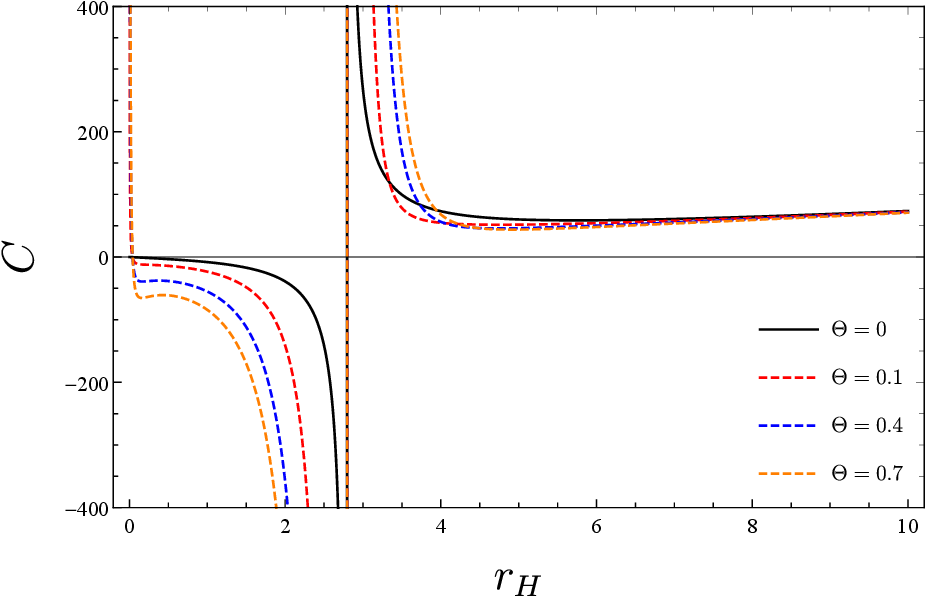}
                \subcaption{$\Lambda=-10^{-3}$}
        \label{fig:heat4}
\end{minipage}
\caption{Heat capacity $C$ as a function of the event horizon radius $r_H$ for $\eta = -1$, with $r_0=1$ and $q=\frac{\pi}{2}$.}
\label{fig:Hheat2}
\end{figure}

The behavior of the heat capacity as a function of the event horizon radius $r_H$ is illustrated in Figures~\ref{fig:Hheat1} and~\ref{fig:Hheat2}. 

For the normal matter case ($\eta=1$), the heat capacity is negative at small horizon radii, indicating local thermodynamic instability. As $r_H$ increases, $C$ crosses zero at a finite value of the horizon radius. This zero defines a \emph{critical horizon} that separates the unstable branch ($C<0$) from the stable branch ($C>0$). Physically, at this point the system no longer exchanges heat with the surroundings, so the transition marks the onset of a thermodynamically stable regime. Beyond the critical horizon, $C$ remains positive and gradually tends toward zero as $r_H$ becomes large, reflecting the vanishing thermal response of very large black holes. The effect of the cosmological constant is illustrated by panels (a) and (b): larger $|\Lambda|$ displaces the critical horizon toward smaller values and broadens the stability region, while smaller $|\Lambda|$ leads to a larger critical horizon and a narrower stable domain.

For the phantom matter case ($\eta=-1$), shown in Fig.~\ref{fig:Hheat2}, the heat capacity displays two distinct singularities. The first, located near $r_H \to 0$, is induced by the NC corrections and is common to both scenarios. The second singularity, however, appears only in the phantom case at a finite horizon radius and originates from the phantom contribution to the geometry. Between these two poles the heat capacity remains negative, signaling persistent instability. Only beyond the phantom–induced pole does $C$ become positive, marking the onset of a stable regime at sufficiently large horizons. As illustrated by panels (a) and (b), larger $|\Lambda|$ sharpen the singularities and broaden the stability domain, while smaller $|\Lambda|$ smooth them and reduce the extent of stability. 

In summary, in both scenarios, the specific heat diverges very close to $r_H \to 0$ as a universal consequence of the noncommutative corrections. The subsequent behavior, however, differs markedly: in the normal case, stability is restored smoothly at the horizon where $C=0$, while in the phantom case, an additional singularity emerges at finite $r_H$, leading to a sharper and more abrupt instability. This contrast shows that phantom matter enriches the thermodynamic phase structure by introducing new singular features beyond the NC corrections.

\begin{figure}[H]
\begin{minipage}[t]{0.5\textwidth}
        \centering
        \includegraphics[width=\textwidth]{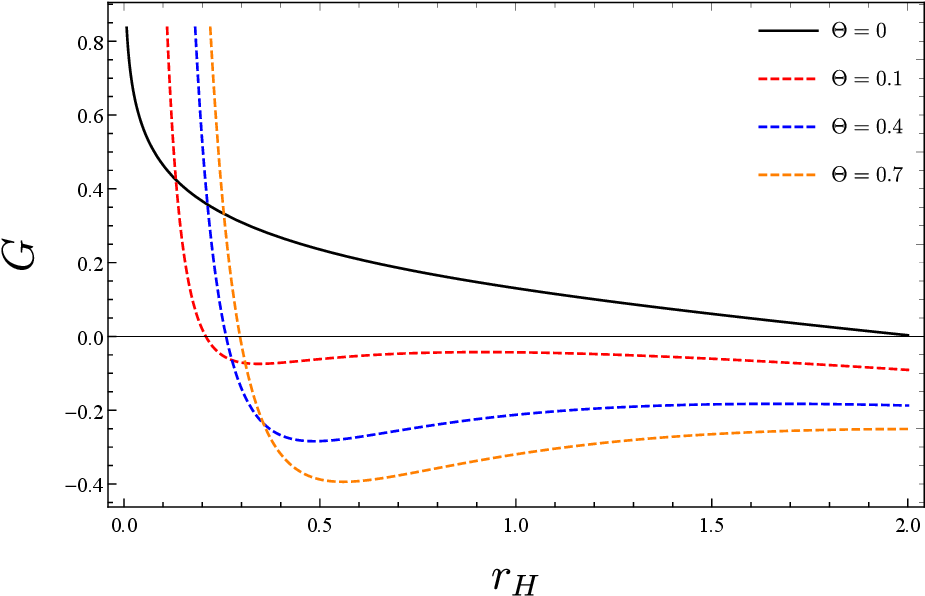}
                \subcaption{$\Lambda=-10^{-2}$}
        \label{fig:fr1}
\end{minipage}
\begin{minipage}[t]{0.5\textwidth}
        \centering
        \includegraphics[width=\textwidth]{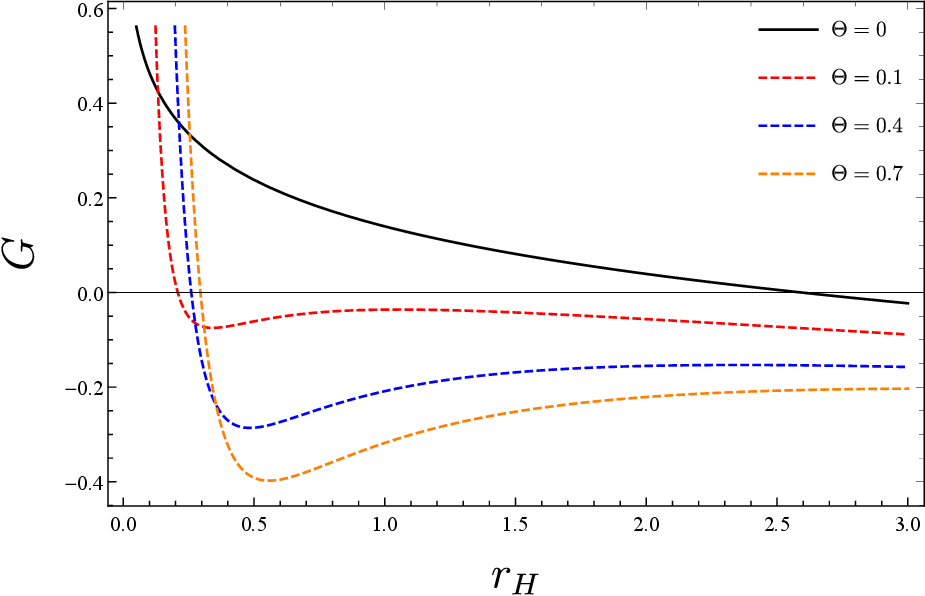}
                \subcaption{$\Lambda=-10^{-3}$}
        \label{fig:fr2}
\end{minipage}
\caption{Gibbs free energy $ G $ as a function of the event horizon radius $ r_H $ for $ \eta = 1 $, with $r_0=1$ and $q=\frac{\pi}{2}$.}
\label{fig:gibbs1}
\end{figure}

\begin{figure}[H]
\begin{minipage}[t]{0.5\textwidth}
        \centering
        \includegraphics[width=\textwidth]{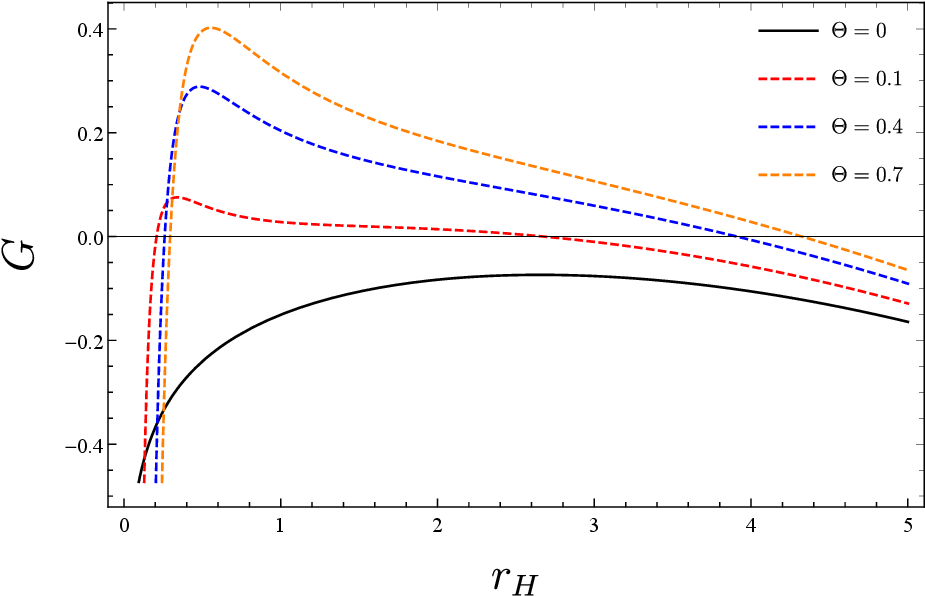}
                \subcaption{$\Lambda=-10^{-2}$}
        \label{fig:fr3}
\end{minipage}
\begin{minipage}[t]{0.5\textwidth}
        \centering
        \includegraphics[width=\textwidth]{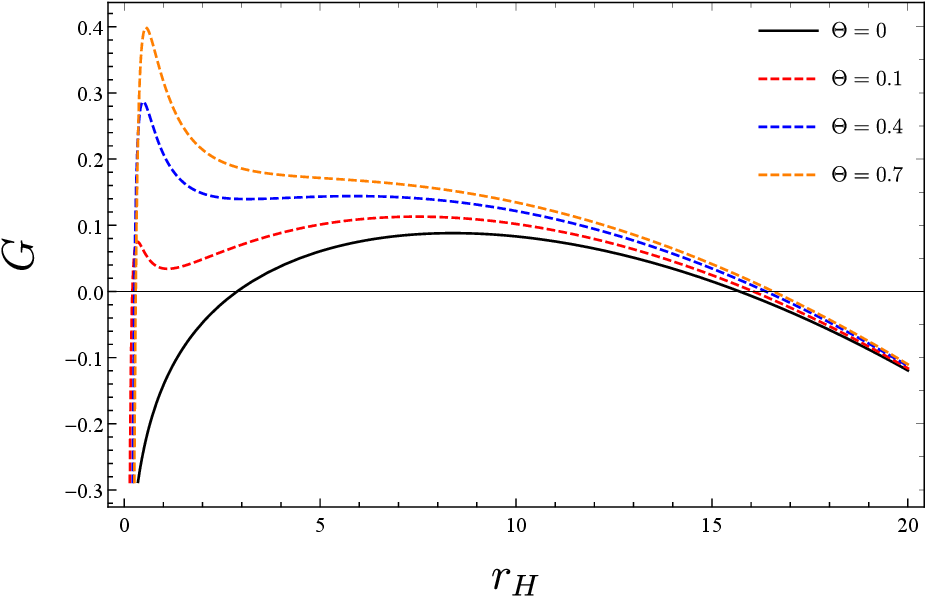}
                \subcaption{$\Lambda=-10^{-3}$}
        \label{fig:fr4}
\end{minipage}
\caption{Gibbs free energy $ G $ as a function of the event horizon radius $ r_H $ for $ \eta = -1 $, with $r_0=1$ and $q=\frac{\pi}{2}$.}
\label{fig:gibbs2}
\end{figure}

To examine the global thermodynamic stability, we now analyze the Gibbs free energy. Within the framework of the canonical ensemble with a fixed charge, the potential, which is also the free energy of the system, presents the thermodynamic behavior of a system in a conventional manner. Using the first law of black hole thermodynamics and interpreting the black hole mass as the black hole enthalpy, the Gibbs free energy of the black hole is given by:
\begin{equation}
G=M-TS. \label{Gibbs}
\end{equation}%
Substituting Eqs.~(\ref{mass}), (\ref{etem}), and (\ref{ent}) in (\ref{Gibbs}), the Gibbs free energy is expressed as:
\begin{eqnarray}
G &=&\frac{\eta q^{2}}{16\pi ^{2}}-\frac{\eta q^{2}}{16\pi ^{2}}\log \Big(\frac{r_{H}}{r_{0}}\Big)+\Lambda r_{H}^{2}-\frac{2\sqrt{\Theta }}{\pi r}\left[ \frac{\eta q^{2}}{8\pi }+\Lambda r_{H}^{2}-\left( \Lambda r_{H}^{2}-\frac{\eta q^{2}}{32\pi ^{2}}\right) \log \Big(\frac{r_{H}}{r_{0}}\Big)\right]   \notag \\
&+&\frac{\Theta }{\pi ^{2}r_{H}^{2}}\left[ \frac{\left( 3\pi ^{2}-8\pi-2\right) }{32\pi ^{2}}\eta q^{2}-\left( \frac{(1+\pi )\eta q^{2}}{8\pi ^{2}}+\Lambda r_{H}^{2}\right) \log \Big(\frac{r_{H}}{r_{0}}\Big)-\frac{\eta q^{2}}{16\pi ^{2}}\log ^{2}\Big(\frac{r_{H}}{r_{0}}\Big)-4\Lambda r_{H}^{2}\right] . \,\, \label{Gibbs1}
\end{eqnarray}
Figures~\ref{fig:gibbs1} and~\ref{fig:gibbs2} illustrate the variation of the Gibbs free energy as a function of the event horizon radius for both normal and phantom cases under different values of the cosmological constant.

For the normal matter case, the Gibbs free energy exhibits characteristics indicative of first-order phase transitions. For smaller black holes 
($G > 0$), the system is thermodynamically unstable, while larger black holes ($G < 0$) achieve stability. The critical radius, where $G = 0$, shifts with the cosmological constant and the noncommutative parameter. A smaller cosmological constant results in a smoother transition and a larger critical radius, while increasing $\Theta $ moves the transition point to higher $r_H $.

In Figure \ref{fig:gibbs2}, for the phantom matter case, the behavior of Gibbs free energy is significantly altered by the exotic properties of phantom energy. The plot reveals a turning point, marked by the condition $\partial G / \partial r_{H}=0$. Panel (a), corresponding to \(\Lambda = -10^{-2}\), shows that increasing $\Theta$ shifts the turning points of $ G $ to larger $ r_H $ values and reduces the magnitude of the local extrema, indicating a weakening of the thermodynamic stability for smaller black holes. In panel (b), with $\Lambda = -10^{-3}$, the variations in $ G $ become smoother, and the impact of $\Theta$ on the stability is less pronounced compared to panel (a). The comparison highlights that a smaller cosmological constant diminishes the effects of non-commutativity while influencing the overall thermodynamic behavior, as reflected in the shifts and smoothness of the curves.

Let us assume that, within the framework of black hole chemistry, the pressure $p$ is related to the cosmological constant through the following relation:
\begin{equation}
p=-\frac{\Lambda}{8\pi} .
\end{equation}
Under this assumption, the black hole mass can be expressed as:
\begin{eqnarray}
M &=&-\frac{\eta q^{2}}{16\pi ^{2}}\log \Big(\frac{r_{H}}{r_{0}}\Big)+\frac{r_{H}^{2}}{8\pi p}-\frac{\sqrt{\Theta }}{\pi r_{H}}\left( \frac{\eta q^{2}}{8\pi }+\frac{\eta q^{2}}{16\pi ^{2}}\log \Big(\frac{r_{H}}{r_{0}}\Big)-\frac{r_{H}^{2}}{8\pi p}\right) \notag \\
&+&\frac{\Theta }{\pi ^{2}r_{H}^{2}}\left( \frac{(\pi -4)\eta q^{2}}{32\pi }+\frac{r_{H}^{2}}{8\pi p}-\frac{\eta q^{2}}{16\pi ^{2}}\log \Big(\frac{r_{H}}{r_{0}}\Big)\right) .  \label{massp}
\end{eqnarray}
The thermodynamic volume $V$, conjugate to the pressure
$p$, is derived from Eq. (\ref{massp}) as:
\begin{equation}
V=\frac{\partial M}{\partial p} =8\pi r_{H} ^{2}  +8\sqrt{\Theta }r_{H}+\frac{8\Theta }{\pi }. 
\end{equation}
The Hawking temperature, previously expressed in Eq.~\eqref{etem}, can now be rewritten as a function of pressure $p$:
\begin{eqnarray}
T &=&8pr_{H}-\frac{\eta q^{2}}{32\pi ^{3}r_{H}}+\frac{\sqrt{\Theta }}{2\pi^{2}r_{H}^{2}}\left[ \frac{\eta q^{2}}{8\pi }-8\pi pr_{H}^{2}+\frac{\eta q^{2}}{16\pi ^{2}}\log \Big(\frac{r_{H}}{r_{0}}\Big)\right]  \notag \\
&+&\frac{\Theta }{2\pi^{3}r_{H}^{3}}\left[ (2-\pi )\frac{\eta q^{2}}{16\pi }-8\pi pr_{H}^{2}+\frac{\eta q^{2}}{16\pi ^{2}}\log \Big(\frac{r_{H}}{r_{0}}\Big)\right].
\end{eqnarray}%
From this, the equation of state for the system is derived as:
\begin{eqnarray}
p &=&\frac{\eta q^{2}}{256\pi ^{3}r_{H}^{2}}+\frac{T}{8r_{H}}+\frac{\sqrt{\Theta }}{16\pi ^{2}r_{H}^{3}}\left[ \pi r_{H}T-\frac{\eta q^{2}}{8\pi }+\frac{\eta q^{2}}{32\pi ^{2}}-\frac{\eta q^{2}}{16\pi ^{2}}\log \Big(\frac{r_{H}}{r_{0}}\Big)\right]   \notag \\
&+&\frac{\Theta }{32\pi ^{3}r_{H}^{4}}\left[ 3\pi r_{H}T+\frac{(4\pi^{2}-12\pi +3)\eta q^{2}}{32\pi ^{2}}-\frac{6\eta q^{2}}{32\pi ^{2}}\log \Big(\frac{r_{H}}{r_{0}}\Big)\right] .  \label{44}
\end{eqnarray}
To determine the critical points, the following conditions must be satisfied:
\begin{equation}
\left. \left( \frac{\partial p}{\partial r_{H}}\right) _{T}\right\vert
_{r_{H}=r_{c}}=0,  \label{48}
\end{equation}%
\begin{equation}
\left. \left( \frac{\partial ^{2}p}{\partial r_{H}^{2}}\right)
_{T}\right\vert _{r_{H}=r_{c}}=0.  \label{49}
\end{equation}
Eqs. \eqref{44} and \eqref{48} represent the conditions for phase coexistence. By solving Eq.~\eqref{48}, the critical temperature $T_{c}$ is obtained as:
\begin{eqnarray}
T_{c} &=&-\frac{\eta q^{2}}{16\pi ^{3}r_{c}}-\frac{\sqrt{\Theta }\eta q^{2}}{64\pi ^{4}r_{c}^{2}}\left[ 1-12\pi -6\log \Big(\frac{r_{c}}{r_{0}}\Big)+\right] + \frac{\Theta \eta q^{2}}{64\pi ^{5}r_{c}^{3}}\left[ 1-8\pi ^{2}+12\pi+6\log \Big(\frac{r_{c}}{r_{0}}\Big)\right].
\end{eqnarray}
Similarly, the critical pressure $p_{c}$ is given by:
\begin{eqnarray}
p_{c} &=&-\frac{\eta q^{2}}{256\pi ^{3}r_{c}^{2}}+\frac{\sqrt{\Theta }\eta q^{2}}{256\pi ^{4}r_{c}^{3}}\left[ 4\pi -1+2\log \Big(\frac{r_{c}}{r_{0}}\Big)\right]+\frac{\Theta \eta q^{2}}{512\pi ^{5}r_{c}^{4}}\left[ 6\log \Big(\frac{r_{c}}{r_{0}}\Big)-6\pi ^{2}+12\pi -1\right].
\end{eqnarray}%
The critical radius $r_{c}$ satisfies the following equation:
\begin{equation}
2\pi ^{2}r_{c}^{2}-6\pi r_{c}\sqrt{\Theta }\left( 2\pi -1+\log \Big(\frac{r_{c}}{r_{0}}\Big)\right) +\Theta \left( 12\pi ^{2}-30\pi +10-15\log \Big(\frac{r_{c}}{r_{0}}\Big)\right) =0.
\end{equation}
Due to the influence of the logarithmic term $\log\left(r_{c}/r_{0}\right)$, it is generally not feasible to determine the critical points analytically. To address this challenge, numerical computations are employed to evaluate the critical thermodynamic quantities for various values of the noncommutative parameter in both the Maxwell and phantom scenarios. The results for the critical thermodynamic quantities under different $\Theta$ values are summarized in Tables \ref{tab:maxwel} and \ref{tab:phantom}.


\begin{table}[H]
\centering
\begin{tabular}{|l|lll|l|}
\hline\hline
\rowcolor{lightgray} 
$\Theta $ & $r_{c}$ & $T_{c}$ & $p_{c}$&$\frac{p_{c} r_{c} }{T_{c} } $ 
\\ \hline\hline

0.3 & 0.390283 & 2.04167$\times 10^{-2}$ & 0.948005$\times 10^{-2}$ & 
\allowbreak 0.181\thinspace 22 \\ 
0.4 & 0.424061 & 1.96779$\times 10^{-2}$ & 0.869447$\times 10^{-2}$ & 
\allowbreak 0.187\thinspace 37 \\ 
0.5 & 0.452517 & 1.85648$\times 10^{-2}$ & 0.806325$\times 10^{-2}$ & 
\allowbreak 0.196\thinspace 54 \\ 
0.6 & 0.477323 & 1.73270$\times 10^{-2}$ & 0.754339$\times 10^{-2}$ & 
\allowbreak 0.207\thinspace 80 \\ 
0.7 & 0.499442 & 1.60607$\times 10^{-2}$ & 0.710545$\times 10^{-2}$ & 
\allowbreak 0.220\thinspace 96 \\ 
0.8 & 0.519487 & 1.48072$\times 10^{-2}$ & 0.672962$\times 10^{-2}$ & 
\allowbreak 0.236\thinspace 10 \\ 
0.9 & 0.537875 & 1.35850$\times 10^{-2}$ & 0.640218$\times 10^{-2}$ & 
\allowbreak 0.253\thinspace 48 \\
\hline\hline
\end{tabular}
\caption{Behavior of the critical parameters $r_{c}$, $T_{c}$, $p_{c}$ and $\frac{p_{c} r_{c}}{T_{c}}$ for different values of the NC parameter in the Maxwell case with $r_0=1$ and $q=\frac{\pi}{2}$.}
\label{tab:maxwel} 
\end{table}

Increasing the noncommutative parameter leads to a larger critical radius $r_c$, a cooling effect with decreasing critical temperature $T_c$, and weaker pressure conditions as indicated by the decline in $p_c$. The ratio $\frac{p_c r_c}{T_c}$ increases consistently, reflecting a steady thermodynamic trend influenced by noncommutative geometry.


\begin{table}[H]
\centering
\begin{tabular}{|l|lll|l|}
\hline\hline
\rowcolor{lightgray} 
$\Theta $ & $r_{c}$ & $T_{c}$ & $p_{c}$&$\frac{p_{c} r_{c} }{T_{c} } $ 
\\ \hline\hline

0.3 & 3.31602 & 0.601952$\times 10^{-2}$ & 0.0730214$\times 10^{-3}$ & 
\allowbreak 0.0402\thinspace 26 \\ 
0.4 & 3.95094 & 0.507873$\times 10^{-2}$ & 0.0518237$\times 10^{-3}$ & 
\allowbreak 0.0403\thinspace 16 \\ 
0.5 & 4.52150 & 0.445444$\times 10^{-2}$ & 0.0397798$\times 10^{-3}$ & 
\allowbreak 0.0403\thinspace 79 \\ 
0.6 & 5.04534 & 0.400336$\times 10^{-2}$ & 0.0320773$\times 10^{-3}$ & 
\allowbreak 0.0404\thinspace 26 \\ 
0.7 & 5.53318 & 0.365876$\times 10^{-2}$ & 0.0267567$\times 10^{-3}$ & 
0.0404\thinspace 64 \\ 
0.8 & 5.99211 & 0.338497$\times 10^{-2}$ & 0.0228761$\times 10^{-3}$ & 
\allowbreak 0.0404\thinspace 96 \\ 
0.9 & 6.42713 & 0.316096$\times 10^{-2}$ & 0.0199293$\times 10^{-3}$ & 
0.0405\thinspace 22 \\
\hline\hline
\end{tabular}
\caption{Behavior of the critical parameters $r_{c}$, $T_{c}$, $p_{c}$ and $\frac{p_{c} r_{c}}{T_{c}}$ for different values of the NC parameter in the phantom case with $r_0=1$ and $q=\frac{\pi}{2}$.}
\label{tab:phantom} 
\end{table}
In the phantom case, the critical radius $r_c$ grows significantly with $\Theta$, while both $T_c$ and $p_c$ decrease more prominently than in the Maxwell case. The ratio $\frac{p_c r_c}{T_c}$ remains nearly constant, showcasing stable thermodynamic behavior despite the extreme effects of phantom energy and noncommutativity.

\section{Black hole as heat engine} \label{sec4}

Now, we analyze the black hole as a thermo-gravitational heat engine.  A heat engine operates in a closed cycle within the $p-V$ plane, absorbing heat energy $Q_H$ from a high-temperature source, converting part of this energy into work $W$, and transferring the remaining heat $Q_C$ to a low-temperature reservoir. The efficiency $\Gamma$ of such an engine is defined as the ratio of the work performed to the heat absorbed, expressed as:
\begin{equation}
\Gamma =\frac{W}{Q_{H}}=1-\frac{Q_{C}}{Q_{H}}.\label{43}
\end{equation}%
The theoretical upper limit for efficiency is achieved by the Carnot engine, which follows an ideal cycle consisting of two isothermal and two adiabatic processes. The efficiency for this idealized cycle is given by:
\begin{equation}
\Gamma =1-\frac{T_{C}}{T_{H}}, \label{444}
\end{equation}
where $T_C$ and $T_H$ denote the temperatures of the cold and hot reservoirs, respectively. 

\begin{figure}[H]
\centering
\includegraphics[scale=0.5]{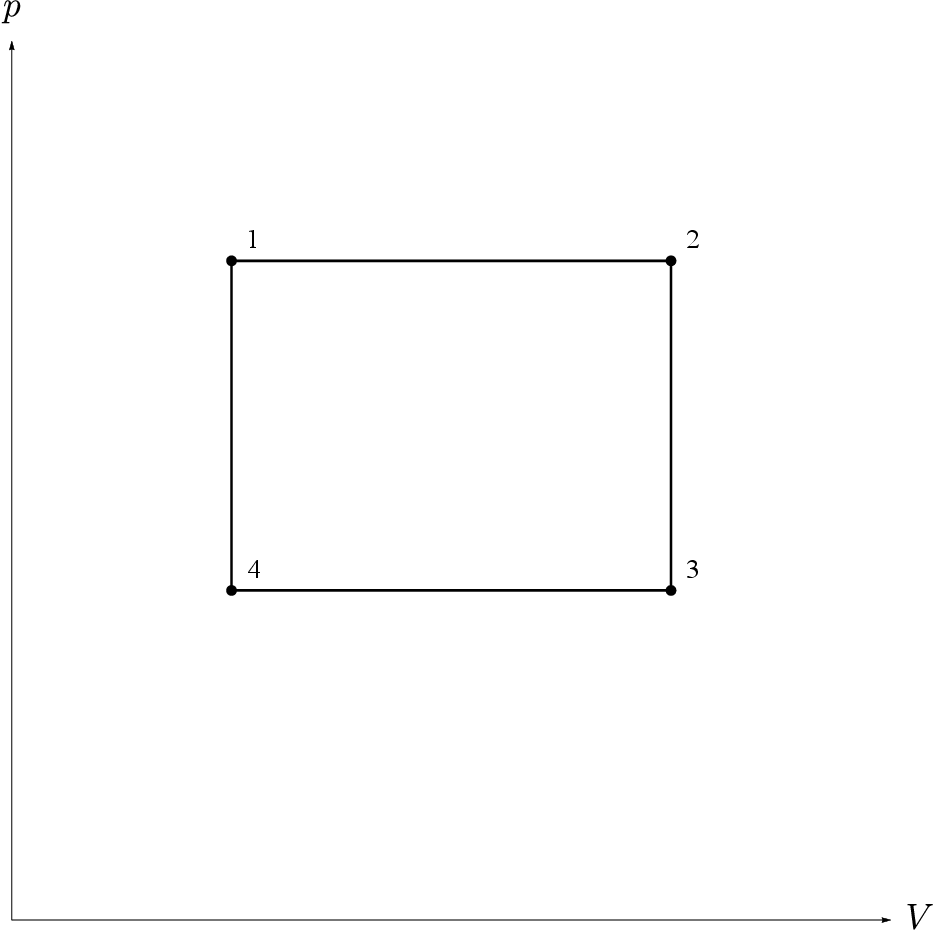}
\caption{Schematic representation of the heat engine cycle in the $P-V$ plane.}
\label{fig:pvfun}
\end{figure}

To model the NC Phantom BTZ black hole as a heat engine, we consider a rectangular thermodynamic cycle in the $P-V$ plane, consisting of two isobaric and two isochoric processes, as illustrated in Figure~\ref{fig:pvfun}.

\noindent The work performed by the heat engine is represented by the area enclosed by the rectangular cycle. For the transitions $1 \to 2$ and $3 \to 4$, the total work done is:
\begin{equation}
W = W_{1 \to 2} + W_{3 \to 4} = p_1 (V_2 - V_1) + p_4 (V_4 - V_3),
\end{equation}
which, using entropy-volume relations for the BTZ black hole, can be rewritten as:
\begin{equation}
W=\frac{8}{\pi }\left( p_{1}-p_{4}\right) \left( S_{2}^{2}-S_{1}^{2}\right) +\frac{8}{\pi }\sqrt{\Theta }\left( p_{1}-p_{4}\right) \left(S_{2}-S_{1}\right). \label{45}
\end{equation}%
For isochoric processes, the heat capacity at constant volume, $C_V$, is zero, implying that no heat exchange occurs during these transitions. Consequently, the heat absorbed, $Q_H$, is determined along the isobaric path $1 \to 2$. The corresponding calculation for $Q_H$ is given by:
\begin{equation}
Q_H = \int_{T_1}^{T_2} C_p dT = M_2 - M_1, \label{47}
\end{equation}
where 
\begin{eqnarray}
M_{2}-M_{1}&=&-\frac{\eta q^{2}}{16\pi ^{2}}\log \Big(\frac{S_{2}}{S_{1}}\Big)+\frac{S_{2}^{2}-S_{1}^{2}}{8\pi ^{3}p_{1}}  \notag \\
&-&\sqrt{\Theta }\left[ \frac{\eta q^{2}}{8\pi }\left( \frac{1}{S_{2}}-\frac{1}{S_{1}}\right) +\frac{\eta q^{2}}{16\pi ^{2}}\left( \frac{1}{S_{2}}\log \Big(\frac{S_{2}}{S_{0}}\Big)-\frac{1}{S_{1}}\log \Big(\frac{S_{1}}{S_{0}}\Big)\right) -\frac{S_{2}-S_{1}}{8\pi ^{3}p_{1}}\right]   \notag \\
&+&\Theta \left[ \frac{(\pi -4)\eta q^{2}}{32\pi }\left( \frac{1}{S_{2}^{2}}-\frac{1}{S_{1}^{2}}\right) -\frac{\eta q^{2}}{16\pi ^{2}}\left( \frac{1}{S_{2}^{2}}\log \Big(\frac{S_{2}}{S_{0}}\Big)-\frac{1}{S_{1}^{2}}\log \Big(\frac{S_{1}}{S_{0}}\Big)\right) \right] .
\end{eqnarray}
Here, $M_1$ and $M_2$ represent the black hole masses at states $1$ and $2$, respectively, and are related to the black hole's thermodynamic properties, including the noncommutative parameter, pressure, and entropy. 

\begin{figure}[H]
\begin{minipage}[t]{0.5\textwidth}
        \centering
        \includegraphics[width=\textwidth]{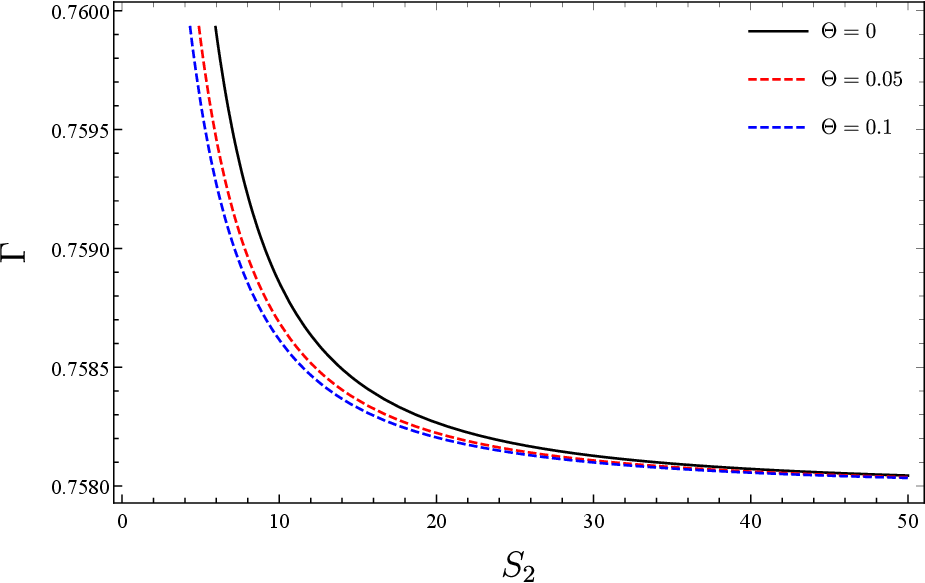}
                \subcaption{$\eta=1$}
        \label{fig:gama1}
\end{minipage}
\begin{minipage}[t]{0.5\textwidth}
        \centering
        \includegraphics[width=\textwidth]{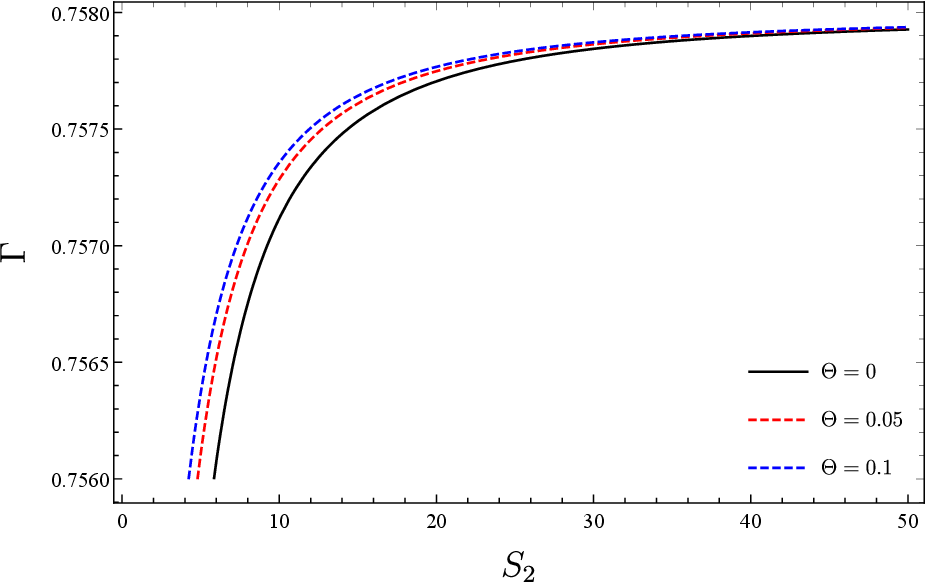}
                \subcaption{$\eta=-1$}
        \label{fig:gamm2}
\end{minipage}
\caption{Plots of the heat engine efficiency $\Gamma$ as a function of entropy $S_2$ for the noncommutative Maxwell ($\eta=1$) and noncommutative phantom ($\eta=-1$) BTZ black holes.}
\label{fig:gamma}
\end{figure}

\begin{figure}[H]
\begin{minipage}[t]{0.5\textwidth}
        \centering
        \includegraphics[width=\textwidth]{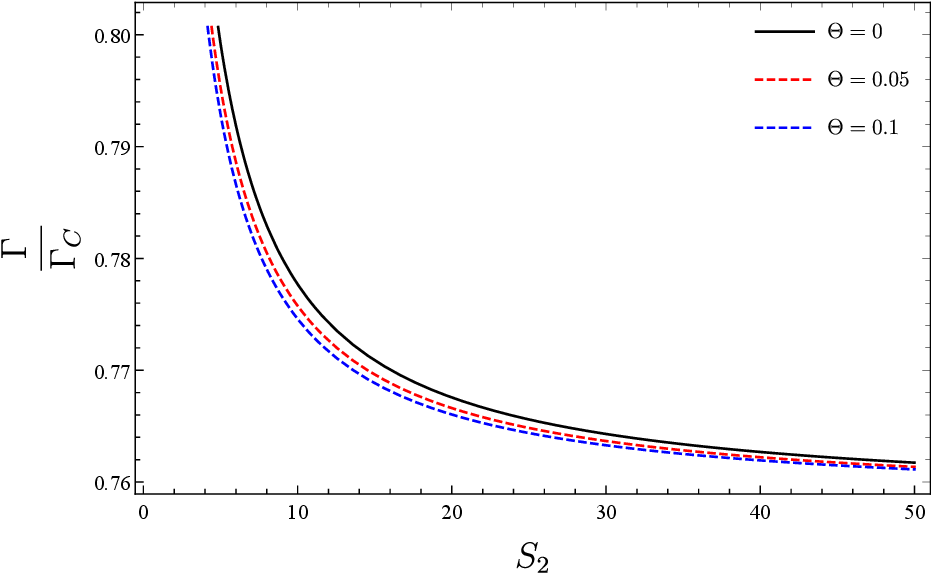}
                \subcaption{$\eta=1$}
        \label{fig:rat1}
\end{minipage}
\begin{minipage}[t]{0.5\textwidth}
        \centering
        \includegraphics[width=\textwidth]{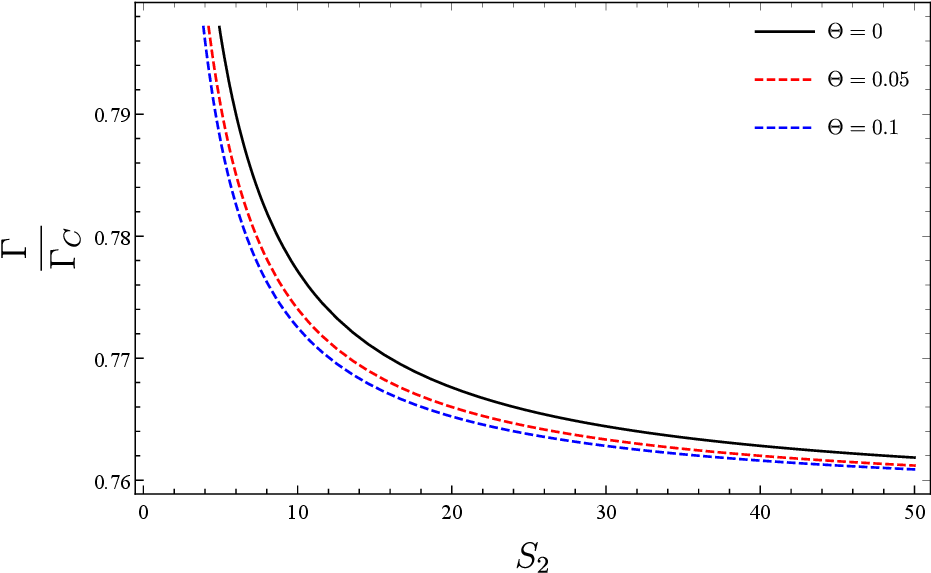}
                \subcaption{$\eta=-1$}
        \label{fig:rat2}
\end{minipage}
\caption{Plots of the ratio ${\Gamma}/{\Gamma_C}$ as a function of entropy $S_2$ for the noncommutative Maxwell ($\eta=1$) and noncommutative phantom ($\eta=-1$) BTZ black holes.}
\label{fig:rat}
\end{figure}

The efficiency of the heat engine, $\Gamma$, can be compared with the efficiency of the Carnot engine, $\Gamma_C$. By associating the higher temperature $T_H$ with $T_2$ and the lower temperature $T_C$ with $T_4$ in Eq.~(\ref{444}), the Carnot efficiency is given as:
\begin{equation}
\Gamma_C = 1 - \frac{T_4(p_4, S_1)}{T_2(p_1, S_2)}.
\end{equation}
Employing Eqs.~(\ref{43}), (\ref{45}), and (\ref{47}), the black hole heat engine efficiency $ \Gamma $ is plotted as a function of entropy $ S_2 $ in Figure~\ref{fig:gamma}, while the ratio $ \Gamma / \Gamma_C $ is depicted as a function of entropy $ S_2 $ in Figure~\ref{fig:rat}.

As shown in Figure~\ref{fig:gamma}, the heat engine efficiency of the noncommutative Maxwell BTZ black hole decreases monotonically with increasing entropy $S_2$ (corresponding to the volume $V_2$) for all values of the noncommutative parameter. This behavior indicates that a larger volume difference between the small black hole $V_1$ and the large black hole $ V_2 $ leads to reduced efficiency. The rate of decrease slows progressively as the volume difference grows, eventually approaching a constant value. Additionally, for a fixed $S_2 $, the efficiency decreases as the noncommutative parameter increases. In contrast, for the noncommutative phantom BTZ black hole, the heat engine efficiency increases monotonically with $ S_2 $, indicating that a larger volume difference between the small black hole and the large black hole enhances the efficiency. Additionally, the efficiency exhibits a rapid initial increase with $ S_2 $, followed by a more gradual growth as the volume becomes larger. For a fixed volume, the efficiency grows with increasing values of $ \Theta $. 

The plots in Figure~\ref{fig:rat} illustrate the ratio of the efficiency $\Gamma$ to the Carnot efficiency $\Gamma_C$ as a function of entropy $S_2$ for noncommutative BTZ black holes under both Maxwell and phantom scenarios. In both cases, it is observed that the efficiency ratio $\Gamma / \Gamma_C$ decreases monotonically with increasing entropy $S_2$, indicating that larger entropies, corresponding to larger horizon areas or volumes, result in reduced thermodynamic efficiency relative to the Carnot limit. The noncommutative parameter introduces a noticeable effect on the efficiency ratio, with larger values of $\Theta$ (e.g., $\Theta = 0.1$) causing a slight suppression of $\Gamma / \Gamma_C$. This behavior is consistent across both Maxwell and phantom cases, though the suppression effect is more pronounced for $\eta = -1$, which is associated with the repulsive gravitational effects of phantom energy. These results highlight the influence of noncommutative geometry and exotic matter fields on the thermodynamic performance of black hole heat engines.

\section{Conclusion} \label{sec5}

In this study, we have examined the thermodynamic and geometric properties of phantom BTZ black holes in a noncommutative (NC) spacetime background. Using Lorentzian-type smeared energy-momentum tensors, we derived a novel modified BTZ metric that captures the combined effects of noncommutativity and exotic phantom fields. This approach offers a technically robust framework that regularizes singularities and incorporates quantum-spacetime features through a geometric lens.

Our investigation revealed several noteworthy results. First, we analytically computed key thermodynamic quantities—Hawking temperature, entropy, and heat capacity—under the influence of NC and phantom corrections. The noncommutative deformation introduces additional terms in the thermodynamic expressions, leading to modified entropy scaling and stability behavior. We identified critical points associated with second-order phase transitions and demonstrated that both local and global thermodynamic stability are sensitive to the NC parameter and phantom coupling.

Second, we formulated and analyzed the behavior of the black hole as a heat engine within the extended phase space formalism. We quantified the mechanical work output and efficiency for cyclic processes, showing that the energy extraction capability is significantly influenced by noncommutative and phantom effects. The introduction of NC geometry leads to nontrivial modulations in the $P$–$V$ diagram and alters the thermodynamic trajectories, thereby refining the standard heat engine framework.

The novelty of this work lies in the combined incorporation of Lorentzian noncommutative geometry and phantom matter within the BTZ black hole background, resulting in new insights into metric structure, phase behavior, and energy conversion. To our knowledge, this is one of the first comprehensive studies to analyze these two effects together in the context of $(2+1)$-dimensional black hole thermodynamics and heat engine efficiency.

Looking ahead, this framework can be generalized to include electric or magnetic charges, rotation, or higher-curvature corrections such as Gauss–Bonnet or Lovelock terms. Moreover, the thermodynamic geometry and holographic dual descriptions of such NC–phantom systems remain open avenues for exploration. Incorporating numerical analysis of critical exponents and exploring analog models may also help connect these theoretical insights with experimentally motivated contexts.

\section*{Acknowledgments}

B. C. L. is grateful to Excellence Project PřF UHK 2205/2025-2026 for the financial support.


\begin{thebibliography}{99}

\bibitem{Hawking1972} S. W. Hawking,  \href{https://doi.org/10.1007/BF01877517}{Commun. Math. Phys. 25, 152–166 (1972).} 

\bibitem{Bekenstein1973} J. Bekenstein, \href{https://doi.org/10.1103/PhysRevD.7.2333?}{Phys. Rev. D \textbf{7}, 2333 (1973).}

\bibitem{Bardeen1973} J. M. Bardeen, B. Carter, S. W. Hawking, \href{ https://doi.org/10.1007/BF01645742}{Commun. Math. Phys. \textbf{31},  161 (1973).}

\bibitem{Hawking1975} S. W. Hawking, \href{https://doi.org/10.1007/BF02345020}{Commun. Math. Phys. \textbf{43}, 199 (1975).}

\bibitem{Banados1992} M. Banados, C. Teitelboim, J. Zanelli, \href{https://doi.org/10.1103/PhysRevLett.69.1849}{Phys. Rev. Lett. \textbf{69}, 1849 (1992).}

\bibitem{Banados1993} M. Banados, M. Henneaux, C. Teitelboim, J. Zanelli, \href{ https://doi.org/10.1103/PhysRevD.48.1506}{Phys. Rev. D \textbf{48}, 1506 (1993).} \emph{Erratum} \href{https://doi.org/10.1103/PhysRevD.88.069902}{Phys. Rev. D \textbf{88}, 069902 (2013).}

\bibitem{Carlip1995} S. Carlip,  \href{https://doi.org/10.1088/0264-9381/12/12/005}{Class. Quantum Grav. \textbf{12}, 2853 (1995).}

\bibitem{Carlip1998} S. Carlip,  \href{https://doi.org/10.1088/0264-9381/15/11/020}{Class. Quantum Grav. \textbf{15}, 3609 (1998).}

\bibitem{Medved2002} A. J. M. Medved,  \href{https://doi.org/10.1088/0264-9381/19/3/313}{Class. Quantum Grav. \textbf{19}, 589 (2002).}

\bibitem{Vagenas2002}
E. C. Vagenas, \href{https://doi.org/10.1016/S0370-2693(02)01695-7}{Phys. Lett. B \textbf{533}, 302 (2002).}

\bibitem{Zhao2006}
R. Zhao, S. L. Zhang,
\href{https://doi.org/10.1016/j.physletb.2006.08.068}{Phys. Lett. B \textbf{641}, 318 (2006).}

\bibitem{Jiang2007} Q. Q. Jiang, S. Q. Wu, X. Cai, \href{https://doi.org/10.1016/j.physletb.2007.05.058}{Phys. Lett. B \textbf{651}, 58 (2007).}

\bibitem{Li2008} R. Li, J. R. Ren, \href{https://doi.org/10.1016/j.physletb.2008.01.077}{Phys. Lett. B \textbf{681}, 370 (2008).}

\bibitem{Gaete2021}
P. Gaete, P. Nicolini, E. Spallucci,   \href{https://doi.org/10.1140/epjc/s10052-021-09301-7}{Eur. Phys. J. C \textbf{81}, 526 (2021).}

\bibitem{Zhao2009} R. Zhao, L. C. Zhang, H. F. Li, \href{https://doi.org/10.7498/aps.58.2193}{Acta Phys. Sin. \textbf{58},  2193 (2009).} 

\bibitem{Iorio2020} A. Iorio, G. Lambiase, P. Pais, F. Scardigli, \href{https://doi.org/10.1103/PhysRevD.101.105002}{Phys. Rev. D \textbf{101}, 105002 (2020).}

\bibitem{Hamil20221} B. Hamil, B. C. L\"{u}tf\"{u}o\u{g}lu, L. Dahbi, \href{https://doi.org/10.1142/S0217751X22501263}{Int. J. Mod. Phys. A \textbf{37}, 2250126  (2022).}

\bibitem{Hamil20222} B. Hamil, B. C. L\"{u}tf\"{u}o\u{g}lu, L. Dahbi, \href{https://doi.org/10.1142/S0217751X22501305}{Int. J. Mod. Phys. A \textbf{37}, 2250130  (2022).}

\bibitem{Hamil20231} B. Hamil, B. C. L\"{u}tf\"{u}o\u{g}lu, L. Dahbi, \href{https://doi.org/10.1142/S0217751X23500525}{Int. J. Mod. Phys. A \textbf{38}, 2350052 (2023).}

\bibitem{Seiberg1999} N. Seiberg, E. Witten, \href{https://doi.org/10.1088/1126-6708/1999/09/032}{J. High Energy Phys. \textbf{09}, 032 (1999).}

\bibitem{Nicolini2005}  P. Nicolini, \href{https://doi.org/10.1088/0305-4470/38/39/L02}{J. Phys. A: Math. Gen. \textbf{38}, L631 (2005).}


\bibitem{Nicolini2006} P. Nicolini,  A. Smailagic, E. Spallucci, \href{https://doi.org/10.1016/j.physletb.2005.11.004}{Phys. Lett. B \textbf{632}, 547 (2006).}

\bibitem{Rizzo2006} T. G. Rizzo, \href{https://doi.org/10.1088/1126-6708/2006/09/021}{J. High Energy Phys. \textbf{09}, 021 (2006).}




\bibitem{Gingrich2010} D. M. Gingrich, \href{https://doi.org/10.1007/JHEP05(2010)022}{J. High Energ. Phys. \textbf{2010}, 22 (2010).}

\bibitem{Ghosh2018} S. G. Ghosh, G. Sushant, \href{https://doi.org/10.1088/1361-6382/aaaead}{Class. Quantum Grav. \textbf{35}, 085008 (2018).}




\bibitem{Anacleto2020} M. A. Anacleto, 
F. A. Brito, J. A. V. Campos, E. Passos, \href{https://doi.org/10.1016/j.physletb.2020.135334}{Phys. Lett. B \textbf{803}, 135334 (2020).}

\bibitem{Ovgun2020} A. Ovgun, I. Sakalli, J. Saavedra, C. Leiva, \href{https://doi.org/10.1142/S0217732320501631}{Mod. Phys. Lett. A \textbf{35}, 2050163 (2020).} 

\bibitem{Hamil} B. Hamil, B. C. L\"{u}tf\"{u}o\u{g}lu, \href{https://doi.org/10.1016/j.dark.2024.101484}{Phys. Dark Universe \textbf{44}, 101484 (2024).}



\bibitem{Dolan2007} B. P. Dolan, K. S. Gupta, A. Stern, \href{https://doi.org/10.1088/0264-9381/24/6/017}{Class. Quantum Grav. \textbf{24}, 1647 (2007).}

\bibitem{Chang2009} E. Chang-Young, L. Daeho, L. Youngone, \href{https://doi.org/10.1088/0264-9381/26/18/185001}{Class. Quantum Grav. \textbf{26}, 185001 (2009).}

\bibitem{Maceda2013} M. Maceda, A. Macias,  \href{https://doi.org/10.1140/epjc/s10052-013-2383-0}{Eur. Phys. J. C \textbf{73}, 2383 (2013).}

\bibitem{Rahaman2013} F. Rahaman, P. K. F. Kuhfittig, B. C. Bhui, M. Rahaman, S. Ray, U. F. Mondal, \href{https://doi.org/10.1103/PhysRevD.87.084014}{Phys. Rev. D \textbf{87}, 084014 (2013).}

\bibitem{Sadeghi2016} J. Sadeghi, V. R. Shajiee, \href{https://doi.org/10.1007/s10773-015-2732-x}{Int. J. Theor. Phys. \textbf{55}, 892 (2016).} 


\bibitem{Anacleto2018} M. A. Anacleto, F. A. Brito, A. G. Cavalcanti, E. Passos, J. Spinelly, \href{https://doi.org/10.1007/s10714-018-2344-x}{Gen. Relativ. Gravit. \textbf{50}, 23 (2018).} 

\bibitem{Gecim2020} G. Gecim, \href{https://doi.org/10.1142/S0217732320502089}{Mod. Phys. Lett. A \textbf{35}, 2050208 (2020).} 

\bibitem{Anacleto2021} M. A. Anacleto, F. A. Brito, B. R. Carvalho, E. Passos, \href{ https://doi.org/10.1155/2021/6633684}{Adv. High Energy Phys. \textbf{2021}, 6633684 (2021).}  

\bibitem{Anacleto2022} M. A. Anacleto, F. A. Brito, E. Passos, J. L. Paulino, A. T. Silva, J. Spinelly, \href{https://doi.org/10.1142/S0217732322502157}{Mod. Phys. Lett. A \textbf{37}, 2250215 (2022).} 

\bibitem{Juric2023} T. Juric, F. Pozar, \href{https://doi.org/10.3390/sym15020417}{Symmetry \textbf{15}, 417 (2023).}

\bibitem{Anacleto2015}  M. A. Anacleto, F. A. Brito, E. Passos, \href{https://doi.org/10.1016/j.physletb.2015.02.056}{Phys. Lett. B \textbf{743}, 184 (2015).}  

\bibitem{Nasseri2005} F. Nasseri, \href{https://doi.org/10.1007/s10714-005-0183-z}{Gen. Relativ. Gravit. \textbf{37}, 2223 (2005).}

\bibitem{Nozari2008} K. Nozari, S. H. Mehdipour, \href{https://doi.org/10.1088/0264-9381/25/17/175015}{Class. Quantum Grav. \textbf{25}, 175015 (2008).}



\bibitem{Ankur2021} Ankur, S. Dey, \href{https://doi.org/10.1016/j.physletb.2021.136391}{Phys. Lett. B \textbf{818}, 136391 (2021).}

\bibitem{Liang2012} J. Liang, B. Liu, \href{https://doi.org/10.1209/0295-5075/100/30001}{EPL \textbf{100}, 30001 (2012).}

\bibitem{Herrera1979} L. Herrera, N. O. Santos, \href{https://doi.org/10.1016/S0370-1573(96)00042-7}{Phys. Rep. \textbf{286}, 53 (1979).}





\bibitem{Bronnikov2006} K. A. Bronnikov, J. C. Fabris, \href{https://doi.org/10.1103/PhysRevLett.96.251101}{Phys. Rev. Lett. \textbf{96}, 251101 (2006).}


\bibitem{Caldwell2002}
R. R. Caldwell, \href{https://doi.org/10.1016/S0370-2693(02)02589-3}{Phys. Lett. B \textbf{545}, 23 (2002).}

\bibitem{Clement2009} G. Clement, J. C. Fabris, M. E. Rodrigues, \href{https://doi.org/10.1103/PhysRevD.79.064021}{Phys. Rev. D \textbf{79}, 064021 (2009).}

\bibitem{Rodrigues2012} M. E. Rodrigues, Z. A. A. Oporto, \href{https://doi.org/10.1103/PhysRevD.85.104022}{Phys. Rev. D \textbf{85}, 104022 (2012).}

\bibitem{Gyulchev2013} G. N. Gyulchev, I. Zh. Stefanov, \href{https://doi.org/10.1103/PhysRevD.87.063005}{Phys. Rev. D \textbf{87}, 063005 (2013).}

\bibitem{Gibbons1996} G. W. Gibbons, D. A. Rasheed, \href{https://doi.org/10.1016/0550-3213(96)00365-3}{Nucl. Phys. B \textbf{476}, 515 (1996).}

\bibitem{Nojiri2003} S. Nojiri, S. D. Odintsov, \href{https://doi.org/10.1016/S0370-2693(03)00594-X}{Phys. Lett. B \textbf{562}, 147 (2003).}

\bibitem{Piazza2004} F. Piazza, S. Tsujikawa, \href{https://doi.org/10.1088/1475-7516/2004/07/004}{JCAP \textbf{07}, 004 (2004).}



\bibitem{Obs1} E. Komatsu et al., \href{http://dx.doi.org/10.1088/0067-0049/192/2/18}{Astrophys. J. Suppl. Ser. \textbf{192}, 18 (2011).}

\bibitem{Obs2} M. Sullivan et al., \href{http://dx.doi.org/10.1088/0004-637X/737/2/102}{Astrophys. J. \textbf{737}, 102 (2011).}

\bibitem{Hannestad2006} S. Hannestad, \href{https://doi.org/10.1142/S0217751X06032885}{Int. J. Mod. Phys. A \textbf{21}, 1938 (2006).}


\bibitem{Bronnikov2012} K. A. Bronnikov,
R. A. Konoplya, A. Zhidenko, \href{https://doi.org/10.1103/PhysRevD.86.024028}{Phys. Rev. D \textbf{86}, 024028 (2012).}

\bibitem{Jardim2012} D. F. Jardim,  M. E. Rodrigues,  S. J. M. Houndjo, \href{https://doi.org/10.1140/epjp/i2012-12123-x}{Eur. Phys. J. Plus \textbf{127}, 123 (2012).} 

\bibitem{Quevdeo2016} H. Quevedo, M. N. Quevedo, A. Sánchez, \href{https://doi.org/10.1103/PhysRevD.94.024057}{Phys. Rev. D  \textbf{94}, 024057 (2016).}

\bibitem{Panah2024} B. Eslam Panah, M. E. Rodrigues, \href{ https://doi.org/10.1140/epjc/s10052-024-13485-z}{Eur. Phys. J. C \textbf{84}, 1125 (2024).}






\bibitem{Johnson2014}  C. V. Johnson, \href{https:dx.doi.org/10.1088/0264-9381/31/20/205002}{Class. Quantum Grav. \textbf{31}, 205002 (2014).}

\bibitem{Johnson2016}  C. V. Johnson, \href{https:dx.doi.org/10.1088/0264-9381/33/21/215009}{Class. Quantum Grav. \textbf{33}, 215009 (2016).}

\bibitem{Hennigar2017} R. A. Hennigar, F. McCarthy, A. Ballon, R. B. Mann, \href{ https:dx.doi.org/10.1088/1361-6382/aa7f0f}{Class. Quantum Grav. \textbf{34}, 175005 (2017).}

\bibitem{Mo2017} J. X. Mo, F. Liang, G. Q. Li, \href{https://doi.org/10.1007/JHEP03(2017)010}{J. High Energy Phys. \textbf{2017}, 10 (2017).}

\bibitem{Balart2019} L. Balart, S. Fernando, \href{https://doi.org/10.1016/j.physletb.2019.07.009}{Phys. Lett. B, \textbf{795}, 638 (2019).}

\bibitem{Liu2017} H. Liu, X.H. Meng, \href{ https://doi.org/10.1140/epjc/s10052-017-5134-9}{Eur. Phys. J. C \textbf{77}, 556 (2017).}

\bibitem{Panah2018} B. Eslam Panah, \href{https://doi.org/10.1016/j.physletb.2018.10.042}{Phys. Lett. B \textbf{787}, 45 (2018).}

\bibitem{Zhang2018} J. Zhang, Y. Li, H. Yu, \href{ https://doi.org/10.1140/epjc/s10052-018-6137-x }{ Eur. Phys. J. C, \textbf{78}, 645 (2018).}

\bibitem{Hendi2018} S. H. Hendi,  B. Eslam Panah, S. Panahiyan, H. Liu, X.-H. Meng,  \href{https://doi.org/10.1016/j.physletb.2018.03.072}{Phys. Lett. B, \textbf{781},  40 (2018).}

\bibitem{Fatima2025} G. Fatima, F. Javed, A. Waseem, B. Almutairi, G. Mustafa, F. Atamuratov, E. G\"udekli, \href{https://doi.org/10.1016/j.dark.2025.101820}{Phys. Dark Universe \textbf{47}, 101820 (2025).}

\bibitem{Mo2018} J. X. Mo, S. Q. Lan, \href{ https://doi.org/10.1140/epjc/s10052-018-6153-x}{Eur. Phys. J. C, \textbf{78}, 666 (2018).}

\bibitem{Chakraborty2019} A. Chakraborty, C. V. Johnson, \href{https://doi.org/10.1142/S0218271819500068}{Int. J. Mod. Phys. D \textbf{28},  1950006 (2019).}

\bibitem{Wei2019}  S. W. Wei, Y. X. Liu, \href{https://doi.org/10.1016/j.nuclphysb.2019.114700}{Nucl. Phys. B \textbf{946}, 114700 (2019).}




\bibitem{Zhang2019} J. Zhang, Y. Li, H. Yu, \href{https://doi.org/10.1007/JHEP02(2019)144}{J. High Energy Phys. \textbf{2019}, 144 (2019).}


\bibitem{Panah2020} B. Eslam Panah, Kh. Jafarzade, S. H. Hendi, \href{ https://doi.org/10.1016/j.nuclphysb.2020.115269}{Nucl. Phys. B, \textbf{961}, 115269 (2020).}


\bibitem{Rajani2020} K. V. Rajani, C. L. A. Rizwan, A. N. Kumara, D. Vaid, K. M. Ajith, \href{https://doi.org/10.1016/j.nuclphysb.2020.115166}{Nucl. Phys. B, \textbf{960},  115166 (2020).}

\bibitem{Ghaffarnejad2020} H. Ghaffarnejad, E. Yaraie, M. Farsam, K. Bamba, \href{https://doi.org/10.1016/j.nuclphysb.2020.114941}{Nucl. Phys. B, \textbf{952}, 114941 (2020).}

\bibitem{Promsiri2021} C. Promsiri, E. Hirunsirisawat, W. Liewrian, \href{ https://doi.org/10.1103/PhysRevD.104.064004}{Phys. Rev. D \textbf{104}, 064004 (2021).}

\bibitem{Roy2021} T. Roy, U. Debnath, \href{https://doi.org/10.1142/S0217751X21501141}{Int. J. Mod. Phys. A \textbf{36}, 2150114 (2021).}



\bibitem{Panah2022}B. Eslam Panah, Kh. Jafarzade, \href{ https://doi.org/10.1007/s10714-022-02904-9}{ Gen. Relativ. Gravit., \textbf{54}, 19 (2022).}



\bibitem{Cao2022}Y. Cao, H. Feng, J. Tao, Y. Xue, \href{https://doi.org/10.1007/s10714-022-02990-9}{Gen. Relativ. Gravit. \textbf{54}, 105 (2022).}


\bibitem{DiMarco2023} M. C. DiMarco, S. L. Jess, R. A. Hennigar, R. B. Mann, \href{ https://doi.org/10.1103/PhysRevD.107.044001}{Phys. Rev. D \textbf{107}, 044001 (2023).}

\bibitem{Nag2023} C. Nag, T. Roy, U. Debnath, \href{https://doi.org/10.1142/S0219887823500937}{Int. J. Geom. Meth. Mod. Phys. \textbf{20}, 2350093 (2023).}



\bibitem{Javed2024} F. Javed, 
G. Fatima, G. Mustafa, S. K. Maurya, B. Almutairi, \href{https://doi.org/10.1016/j.dark.2024.101677}{Phys. Dark Universe \textbf{46}, 101677 (2024).}

\bibitem{Kruglov2024} S. I. Kruglov, \href{https://doi.org/10.1139/cjp-2024-0146}{Canadian J. Phys. \textbf{103}, 448 (2024).}

\bibitem{Panah2025} B. Eslam Panah, N. Heidari, \href{https://doi.org/10.1016/j.jheap.2024.12.001}{J. High Energy Astrop. \textbf{45}, 181 (2025).}

\bibitem{Mondal2025} D. Mondal, U. Debnath, A. Pradhan, \href{https://doi.org/10.1142/S0219887825300028}{Int. J. Geom. Meth. Mod. Phys. \textbf{xx}, xx (2025). https://doi.org/10.1142/S0219887825300028}


\end{thebibliography}
\end{document}